\documentclass{article}

\usepackage{arxiv}

\usepackage[utf8]{inputenc} 
\usepackage[T1]{fontenc}    
\usepackage{hyperref}       
\usepackage{url}            
\usepackage{booktabs}       
\usepackage{amsfonts}       
\usepackage{nicefrac}       
\usepackage{microtype}      
\usepackage{lipsum}

\usepackage{times}
\usepackage{epsfig}
\usepackage{graphicx}
\usepackage{amsmath}
\usepackage{amssymb}
\usepackage{multirow}
\usepackage{subfigure}
\usepackage{amsthm,amsmath,amssymb}
\usepackage{mathrsfs}
\usepackage{paralist}

\title{Perceptual Image Restoration with High-Quality Priori and Degradation Learning}

\author{
  Chaoyi Han \thanks{hancy16@mails.tsinghua.edu.cn}\\
\And
 Yiping Duan \\
\And
Xiaoming Tao \\
\And
Jianhua Lu \\
}

\begin{document}
\maketitle

\begin{abstract}

Perceptual image restoration seeks for high-fidelity images that  most likely degrade to given images.   
For better visual quality, previous work proposed to search for solutions within the natural image manifold, by exploiting the latent space of a generative model.
However, the quality of generated images are only guaranteed  when latent embedding lies close to the prior distribution.
In this work, we propose to restrict the feasible region within the prior manifold. 
This is accomplished with a non-parametric metric for two distributions: the Maximum Mean Discrepancy (MMD). 
Moreover, we model the degradation process directly as a conditional distribution. We show that our model performs well in measuring the similarity between restored and degraded images.  
Instead of optimizing the long criticized pixel-wise distance over degraded images, we rely on 
such model to find visual pleasing images with high probability.  
Our simultaneous restoration and enhancement framework generalizes well to real-world complicated degradation types.
The experimental results on perceptual quality and no-reference image quality  assessment (NR-IQA)  demonstrate the superior performance of our method. 

\end{abstract}

\section{Introduction}

Image restoration is a longstanding problem in image processing and computer vision. 
The need of restoration is deeply rooted among acquisition, transmission and application related to images. We call the process from high-quality to low-quality images as degradation, the inverse to restoration. Restoration method is highly related to corresponding degradation process.    
According to specific model, it can be further divided into deblurring, denoising and super-resolution, to name a few. 

Recent advances in deep learning have greatly improved the state-of-the-art performances in those subareas \cite{Tao2018ScaleRecurrentNF,Zhang2019DeepSH,Zhang2017BeyondAG,Zhang2020MemoryEfficientHN}.
However, most previous work reduces to an end-to-end trained neural network with pixel-wise $\ell_2$ loss.
The pixel-wise loss has the well-known drawbacks of leading to fuzzy in high-variance areas. In other words, the $\ell_2$ norm is not so in line with perceptual quality. On consider that in most cases, the ultimate target of image restoration is a better perceived quality from human, the pixel-wise distance is not a perfect choice.  Recent work has reconsidered this topic from a perceptual-oriented framework.
Most of them leverage the advances in generative models like Generative Adversarial Networks (GAN) \cite{GAN}, or perceptual similarity metrics \cite{Dosovitskiy2016GeneratingIW} to get a close approximation to subjective experience \cite{Ledig2017PhotoRealisticSI,Yang2020FromFT}.
They adopt a combination of $\ell_2$ and perceptual loss when training models, thus leading to a tradeoff between PNSR and perceptual quality.
Despite their improvement, the fashion of simply putting the pieces together is wired. 
The $\ell_2$ loss still involves in the training process and hampers the restored images from looking realistic.

Some researchers have attempted to reformulate the problem to exclude the pixel-wise loss when generating high-quality images.
They proposed to take the output of generative models as restored images, since this ensures the output quality with a well trained generative model \cite{PCGAN,stylegan,stylegan2}.
In this case, the pixel-wise loss (or perceptual loss) is used only for searching embedding in the latent space of corresponding generative model. 
Previous work has exploited the possibility and ways of embedding images into the latent space of GANs \cite{Abdal2019Image2StyleGANHT,Bau2019SeeingWA,Gu2020ImagePU}.

Although the transformation from original image space to latent space makes it easier to generate images within the natural image manifold, one key issue remains. 
Most generative models start with a specific distribution. Once trained, the visual quality of generated images are  guaranteed only when the corresponding latent vector lies in area with high probability. Again, the problem remains how to restrict the feasible region within given manifold. \cite{2020PULSE} proposed a spherical optimization method to deal with Gaussian prior and an approximation of the mapping network in StyleGAN \cite{stylegan}. In \cite{stylegan2}, they renormalized the signals to meet the prior distribution after each optimization step.  
However, those heuristic methods are indeed hand-crafted and apply to simple prior distribution. 

For an arbitrary prior distribution in latent space, the only and important information comes from samples. We propose to guide the optimization process with the sampled statistics. We treat the  embedding to be solved as samples from a specific distribution, and aim to minimize the distance between such distribution and the prior distribution. 
We adopt the Maximum Mean Discrepancy (MMD) distance \cite{Gretton2012AKT} for this task, and explicitly incorporate it into the optimization objective using the empirical formulation with Gaussian kernel. The method is parameter-free and does not involve extra models to capture the prior distribution. We find such constraint very effective and bring substantial improvement on the perceptual quality. 

Another issue is the optimization objective used in the restoration process.
For restoration, we should find the images degrade correctly. However, real-world degradation is complicated. The simple additive noise model and downscale model are far from enough. 
Previous work \cite{2018To,Zhao2018UnsupervisedDL} attempts to learn a degradation model from real-world image pairs.  But only the learned mapping from high-quality to low-quality images is incomplete. We would still have to the adopt $\ell_{p}$ distance between the input degraded image and that from the degradation model as the measurement of how "correct" the restored images are.  
However, the degradation model is trained independently. There is no guarantee from design that a degradation model followed by the $\ell_{p}$ distance is a good indicator of correctness.

Besides the "degradation mapping+ $\ell_{p}$ distance" fashion, we propose a different framework to 
accomplish this in one step. 
We directly model the degradation process as the conditional distribution, namely $p(degraded|original)$. Such model explicitly gives the probability from restored image to the given degraded image. It naturally generalizes to situations where the degradation degree is unknown, i.e. the original image could degrade to a set of images. The "degradation mapping" used in \cite{2018To,Zhao2018UnsupervisedDL} does not consider such situation and assumes a "one-to-one" mapping. 
The overall framework is shown in Figure \ref{framework}.

Our  contributions are as follows.

\begin{enumerate}
\setlength{\parsep}{0ex}
    \item We propose a novel optimization strategy for restoring images with the GAN prior. To fit the prior distribution, previous work  has used hand-crafted and simple methods. The proposed method is parameter-free and applies to arbitrary prior distribution in latent space. We show that our strategy greatly improves the perceptual quality of restored images.
    \item We formulate a universal degradation model in the form of conditional distribution. The model could explicitly tell the correctness of the restoration and guide the optimization process. 
    \item With the above two modules, we actually present a novel framework for perceptual image restoration.
\end{enumerate}

\section{Related Work}

There are much work regarding image restoration and generative models. We briefly review several work that is most related to our method.

\noindent \textbf{Perceptual Image Restoration}. Image restoration aims to reverse the degradation process and find the original high-quality images. However, this is in most cases an ill-posed problem \cite{Rani2016ABR}. Taking image super-resolution as example, there are lots of images that could downscale to the same low-resolution image. Therefore, a simultaneous enhancement is desired during restoration. In other words, previous work wants to find the possible images as well as remain  high-quality \cite{Blau20182018PC}. Most of them adopt perceptual metric \cite{Dosovitskiy2016GeneratingIW} ( such as VGG\cite{VGG} feature matching loss) and generative models \cite{GAN,pixelcnn,pixelcnn++,Dinh2017DensityEU} to accomplish this.
Among them GAN is the most widely used model for its superior performance  in generating high-quality images. 
Furthermore, recent work has proposed to take the output from GAN as desired result for common image processing problems \cite{2020PULSE,Gu2020ImagePU}. Such framework has achieved great improvement on the subjective quality. The key insight is excluding any reference-based metrics from resotration process to best preserve the ability of generative models. In consequence, they count on the generality and degradation model to ensure correctness, which is the inherent defect from design. Besides, they have to  deal with the generative model carefully to find a proper embedding.

\noindent \textbf{Editing Image with GAN prior}. The goal of image restoration and enhancement is transforming low-quality images into high-quality images. However, it is essential to known in advance what high-quality actually means.  Early work treat high-quality as simple statistics like color, contrast and brightness. Recent advances on generative models learn such statistics from real images. 
With a well trained GAN, \cite{Abdal2019Image2StyleGANHT} find that we can embedding images into the latent space. This provides the way of using GANs as the high-quality prior and \cite{Gu2020ImagePU} generalizes the GAN prior to common image processing problems. To embedd images into the latent space, previous work either directly optimizes the latent codes or trains a reverse encoder \cite{Zhu2016GenerativeVM}. The former method requires no addiontal training and becomes the major choice \cite{Abdal2019Image2StyleGANHT,Shen2020InterpretingTL,stylegan2,2020PULSE,Gu2020ImagePU}. 

\noindent \textbf{Image Degradation Model}. Some subareas of image restoration have rather simple degradation model, like additive noise and downscale, while some are not, like compression artifacts of JPEG \cite{JPEG}, JPEG2000 \cite{Adams2000JPEG2} and learning based compression \cite{Toderici2017FullRI,Han2020TowardVG}. 
Although there are many learning based methods advancing the performance in those subareas, they still assume a simple degradation pattern \cite{Zhang2019DeepSH,Zhang2017BeyondAG}. The simple assumption cannot handle many complicated practical degradations, which are more concerned in real world applications. Therefore, some researchers proposed to learn to degradation model from real world images \cite{2018To,Zhao2018UnsupervisedDL}. They adopt GANs to learn to degradation mapping, and then train corresponding restoration models with generated images.

\section{Method}
We start with the framework of the proposed perceptual image restoration.
Let $I_H \in \mathbb{R}^{M\times N}$ denotes the original high-quality image and $I_L \in \mathbb{R}^{m\times n}$ the degraded image produced through  arbitrary fashion $\mathcal{F}$,
\begin{equation}   
\mathcal{F}: I_H \to I_L
\end{equation}
The restoration from degraded image $I_L$ seeks for the pseudo-inverse model $\mathcal{G}$ that gives $\mathcal{G}(I_L) \in \{I_H|\mathcal{F}(I_H)=I_L\}$. The inverse model could be ill-posed and hence we also expected the restored images lies in the natural image manifold  $\mathcal{M}$ for sake of realistic,
\begin{equation}
\mathcal{G}(I_L) \in \{I_H|\mathcal{F}(I_H)=I_L\} \cap \mathcal{M}
\end{equation}

In many cases, such degradation model may not be deterministic. For example, a Guassian blurring model would involve stochastic noise. Therefore, a more general degradation model comes as,
\begin{equation}   
\mathcal{F}: I_H \to \{I_L|I_L \sim P_{I|I_H}\}
\end{equation}
where $P_{I|I_H}$ denotes the probability distribution of all possible images degraded from $I_H$.
The set of degraded images form a manifold, in which images would share similar content with the original image. We assume such correspondence and consistence can be fully captured  by the conditional distribution $P_{I|I_H}$, which is uniquely determined by the degradation model. 
The degradation manifold for a specific image can be blurred images with Gaussian noise, or compressed images with unknown ratios, to name a few. 

Similarly, we still expect the restored images to be natural. We seek for images with high probability within the natural image manifold,
\begin{equation}   
\mathcal{G}: I_L \to \{I_H | P_{I_L|I_H} > \epsilon \} \cap \mathcal{M}  
\end{equation}
Till now we have introduced the overall framework of our perceptual image restoration.

\begin{figure*}
\centering
\includegraphics[width=1\linewidth]{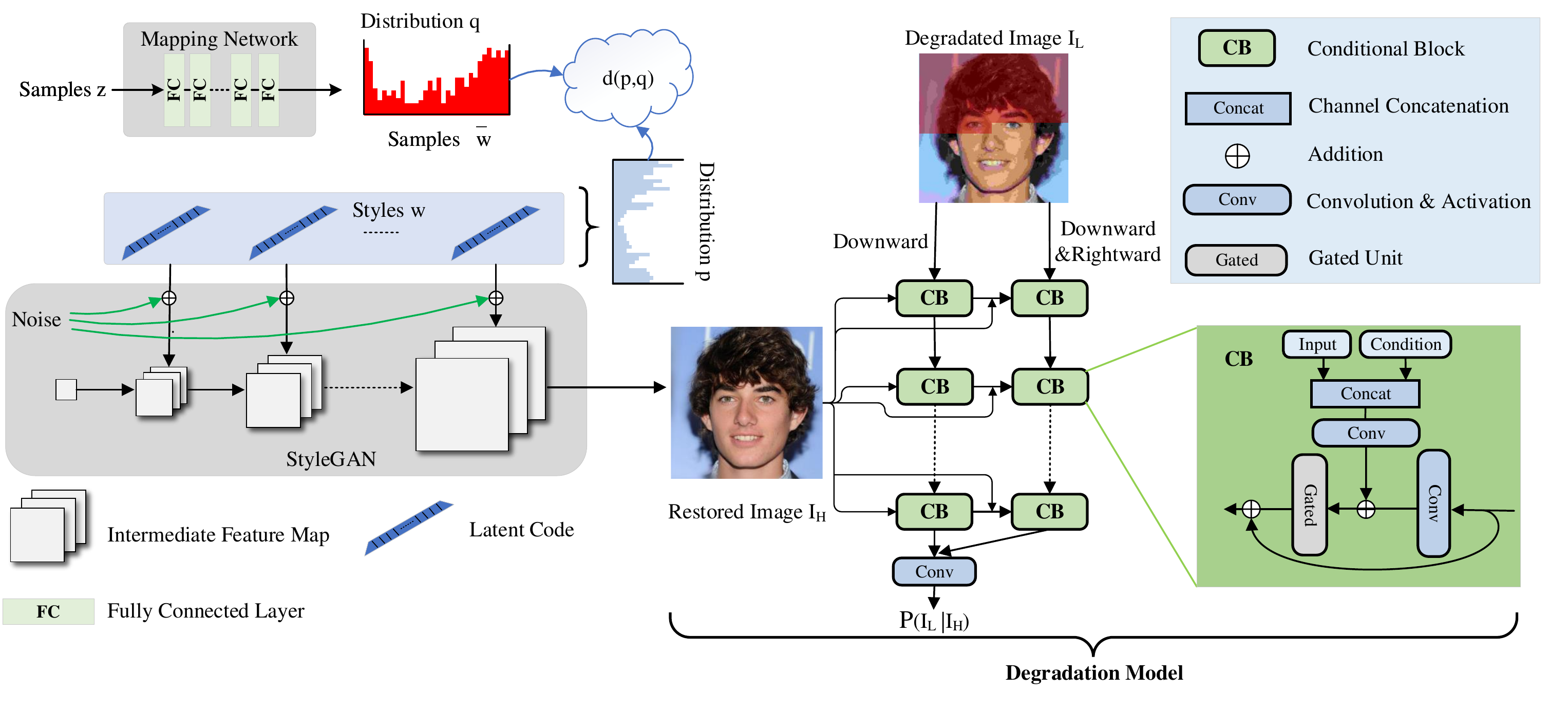}\vspace{4pt}
\caption{The framework of the proposed method. The pipeline is divided into two parts, namely the generative model (StyleGAN) and the degradation model. We optimize the latent codes in $\mathcal{W}$ space, also called styles. During optimization, we explicitly force the styles stay close to the prior distribution. The  statistics of prior distribution is captured with samples generated from the mapping network. The degradation model could tell the correctness of restored image using the probability $P(degraded|restored)$. With the above modules together, we are able to restore the most possible images while retain natural and high-quality.}
\label{framework}
\end{figure*}

\subsection{Explicit Constraint on Latent Space Manifold }
When it comes to an accessible model that constrained within the natural image manifold, the generative model is mostly referred. There has been much work exploring the latent space of a generative model and we also get start with those work. Typically, a pretrained generative model is supposed to output a high-quality natural image  from a latent vector $g: z \to I_H$. Here, $z \in \mathbb{R}^d$ denotes the d-dimensional latent embedding. 
For the restored image to fall into natural image manifold $\mathcal{M}$, most work tend to find a latent vector $\hat{z}$ and adopt $g(\hat{z})$ as desired output. 
Recent advances show that, by leveraging a generative model, we are possible to find restored images with much better visual quality.
However, as noted before, the transformation from original image space to latent space does not fully bypass the obstacle.
It is general knowledge that the visual quality of generated images are  guaranteed only when the corresponding latent embedding lies in area with high probability. 

Again, the problem remains how to restrict the feasible region within given manifold. This is a non-trival problem, even when the explicit formulation of the prior distribution is known. As noted in  \cite{2020PULSE,Bora2017CompressedSU}, the direct likelihood loss always force the latent embedding to the point with highest probability, instead of a feasible region. 
\cite{2020PULSE} proposed to restrict the searching space within a sphere of radius $\sqrt{d}$ for high-dimensional Gaussian, near which lies most of the mass. However, this is a rather rough and handcrafted solution, which is applicable only to Gaussian prior, not even the $\mathcal{W} \subseteq \mathbb{R}^d$ space in StyleGAN \cite{stylegan,stylegan2}. As our known, no previous work has ever given a proper constraint on the searching process in $\mathcal{W}$ space.

We start with the StyleGAN model  with the $\mathcal{W}$ space as feasible region.
As pointed out by \cite{Shen2020InterpretingTL,Abdal2019Image2StyleGANHT}, $\mathcal{W}$ space achieves much better performance over $\mathcal{Z}$ space for  embedding images. 
Specifically, the $\mathcal{W}$ space consists of several latent vectors $\{w_1,w_2,...,w_k\}$ that defines the styles of output image. 
In original StyleGAN, the latent vectors are identical $w_1 \equiv w_2 \equiv... \equiv w_k$. For generalization to unseen images, they are allowed to change separately. But we still expect them fall into the same manifold. 
We treat $\{w_1,w_2,...,w_k\}$ as independent samples from a distribution $q_\mathcal{W}: \mathbb{R}^d \to R, \int_{\mathbb{R}^d} q_\mathcal{W}=1  $. Let $p_\mathcal{W}$ denotes the prior distribution. We expect $q_\mathcal{W}$ lie as close to $p_\mathcal{W}$ as possible,
\begin{equation}   
d(q_\mathcal{W},p_\mathcal{W}) < \epsilon
\end{equation}
This is achieved with the Maximum Mean Discrepancy (MMD) \cite{Gretton2012AKT},  a non-parametric distance metric for two distributions,   
\begin{equation} 
\begin{aligned}
\text{d}^{2}  \left[k,  p_\mathcal{W}, q_\mathcal{W}  \right] = & \mathop{\mathbb{E}}\limits_{x,x'\sim p_\mathcal{W}} \left[k(x,x')\right]  - 2\mathop{\mathbb{E}}\limits_{x\sim p_\mathcal{W},y\sim q_\mathcal{W}} \left[k(x,y) \right] \\ 
& + \mathop{\mathbb{E}}\limits_{y,y'\sim q_\mathcal{W}} \left[k(y,y') \right]
\end{aligned}
\end{equation}
where $k$ is the kernel function. Such non-parametric metric avoids  extra training efforts. To achieve this, we take several samples from the Gaussian prior distribution and forward them through the mapping network to obtain samples in the  $\mathcal{W}$ space, namely $\{\bar{w_1},\bar{w_2},...,\bar{w_{k'}}\}$. With the Gaussian kernel, the distance turns into,
\begin{equation} 
\begin{aligned}
&\text{d}^{2}  \left[\gamma,  p_\mathcal{W}, q_\mathcal{W}  \right] =\frac{1}{k^2} \sum_{1\le i,j \le k}e^{-\frac{||w_i- w_j||_2}{\gamma}} + \\
&\frac{1}{k'^2}\sum_{1\le i,j \le k'}e^{-\frac{||\bar{w_i}- \bar{w_j}||_2}{\gamma}} 
 - \frac{2}{kk'} \sum_{1\le i \le k,1 \le j\le k'}e^{-\frac{||w_i-\bar{w_j}||_2}{\gamma}} 
\end{aligned}
\end{equation}
where $\gamma$ denotes the bandwidth of the Gaussian kernel. We empirically select the bandwidth according to the sampling statistics in $\mathcal{W}$ space. 

\subsection{Degradation Estimation}
Another essential part for image restoration is the degradation model. Some degradation models are straightforward (like additive noise) while others are not. As note in \cite{2020PULSE,2018To,Zhao2018UnsupervisedDL}, real-world degradation models are complicated and lack explicit formulation. 
Previous work \cite{2018To,Zhao2018UnsupervisedDL} attempts to learn a degradation model using generative adversarial networks (GAN) from real-world images. However, a degradation model from high-quality images to low-quality images are not enough. What we actually need is kind of indicator on how possible the restored images could degrade to the input images. In \cite{2020PULSE}, this is accomplished with the $l_2$ norm  under the simple downscale model. The $l_2$ norm is direct but makes no sense with a learned degradation model, because the smaller $l_p$ distance between degraded images does not guarantee a smaller distance between original images (objective or subjective). 

We model the degradation process $\mathcal{F}: I_H \to \{I_L|I_L \sim P_{I|I_H}\}$ directly as a density estimation problem, and adopt a conditional version of autoregressive model \cite{pixelcnn,pixelcnn++} to accomplish this subtask. The illustration is shown in Figure \ref{framework}. We adopt the pixelcnn++ framework \cite{pixelcnn++}, but with a rather small receptive field. 
We keep the gated unit and the discretized mixture of univariate Gaussian model used in \cite{pixelcnn++}, but discard the downsampling and upsampling  branch.
An accurate estimation of high-dimensional distribution requires numerous of samples.
As pointed out in \cite{MultivariateDE}, the growth in sample size is at least exponential with dimension in order to attain an equivalent amount of accuracy in terms of mean integrated squared error (MISE). For images with size $256 \times 256$, corresponding sample size would be an enormous number ($e^{10^5}$ in exponential). As a result, we only model the local degradation which performs better with limited samples. Otherwise the model would suffer from severe overfiting, as previously noted in \cite{pixelcnn,pixelcnn++}.
We would emphasize that the local design is not troublsome, since image restoration is usually a low-level problem and does not involve much global information.
Lots of degradations such as noise, blur, and JPEG (block based) are local from origin. We provide the condition (high-quality image) for each "Conditional Block" in  Figure \ref{framework}, so that the final output has direct access to local pixel information as well as a larger context from previous layers.

\section{Experiments}

\subsection{Settings}

As noted in \cite{stylegan2}, the original StyleGAN model suffers from unneglectable artifacts. Therefore, we replace StyleGAN with the improved version in \cite{stylegan2}  as backbone model.
Following previous work, we test out method in the CelebA HQ dataset \cite{celeba-hq} with 100 random samples. 
We keep the spherical gradient descent and the cross loss proposed in \cite{2020PULSE}. Differently, we do not set the threshold for convergence and run 100 steps for all test images. Besides, we optimize the latent code in $\mathcal{W}$ space directly instead of the $\mathcal{Z}$ space  as in \cite{2020PULSE}.
The optimization initializes from the mean embedding  of $\mathcal{W}$ space $\mathbb{E}_{z \sim P(z)}[f(z)]$,  the same strategy used in \cite{Abdal2019Image2StyleGANHT,stylegan2}.
Previous study indicates that such initialization works reliably for face images.
As for the MMD loss, we take 1000 samples in the $\mathcal{W}$ space using the mapping network. We set the bandwidth $\gamma$ to 512. The degradation model contains 6 conditional blocks after the downward and rightward stream with channel number set to 100.

\begin{figure}
\centering
\subfigure[LR]{
\begin{minipage}[b]{0.2\linewidth}
\includegraphics[width=1\linewidth]{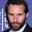}\vspace{4pt}
\includegraphics[width=1\linewidth]{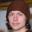}\vspace{4pt}
\includegraphics[width=1\linewidth]{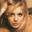}\vspace{4pt}
\includegraphics[width=1\linewidth]{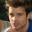}\vspace{4pt}
\end{minipage}}
\subfigure[BICUBIC]{
\begin{minipage}[b]{0.2\linewidth}
\includegraphics[width=1\linewidth]{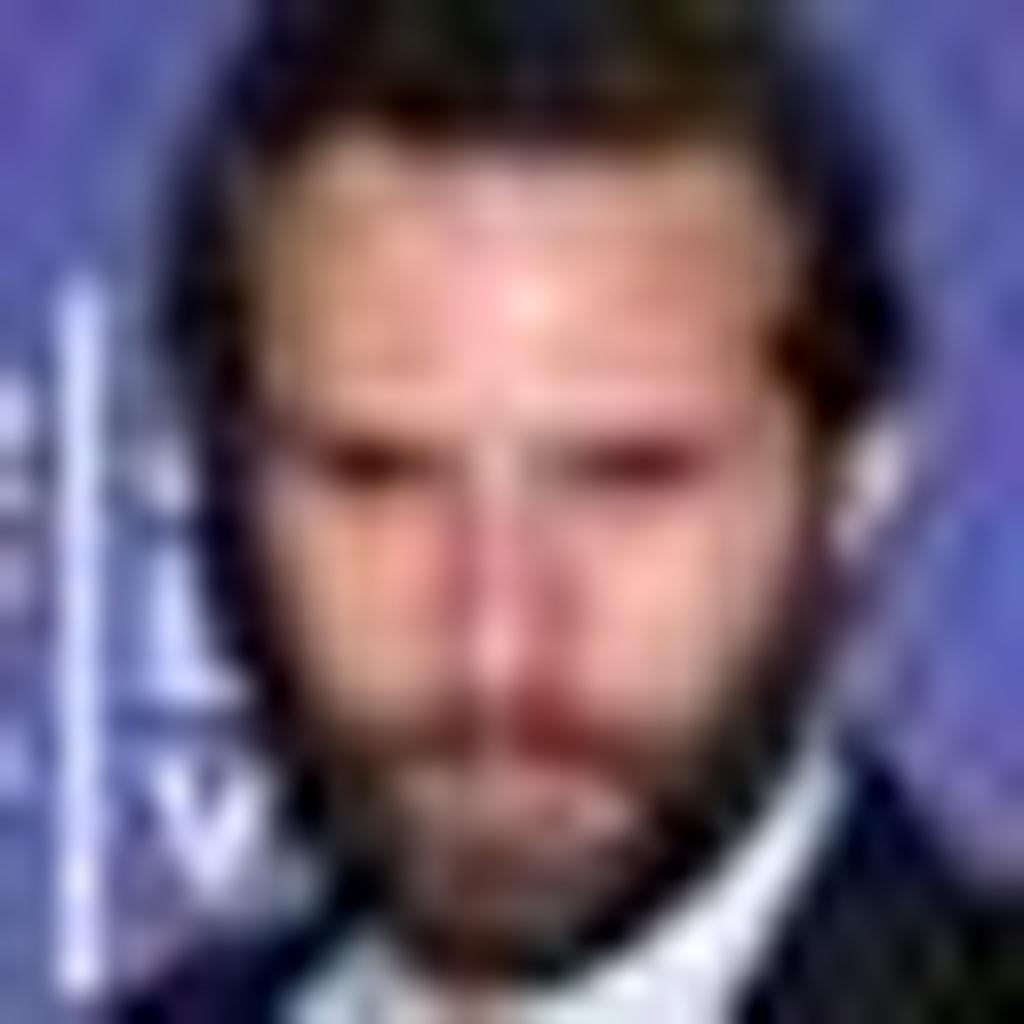}\vspace{4pt}
\includegraphics[width=1\linewidth]{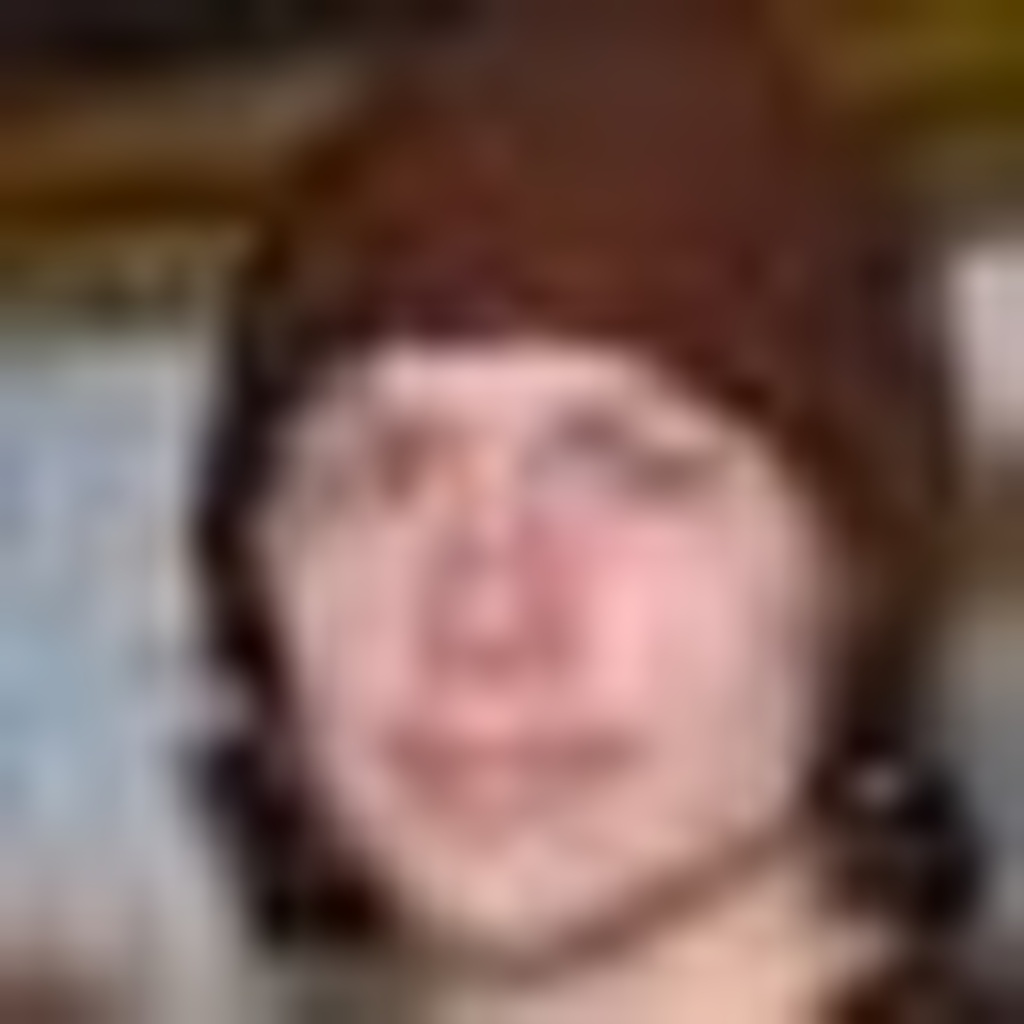}\vspace{4pt}
\includegraphics[width=1\linewidth]{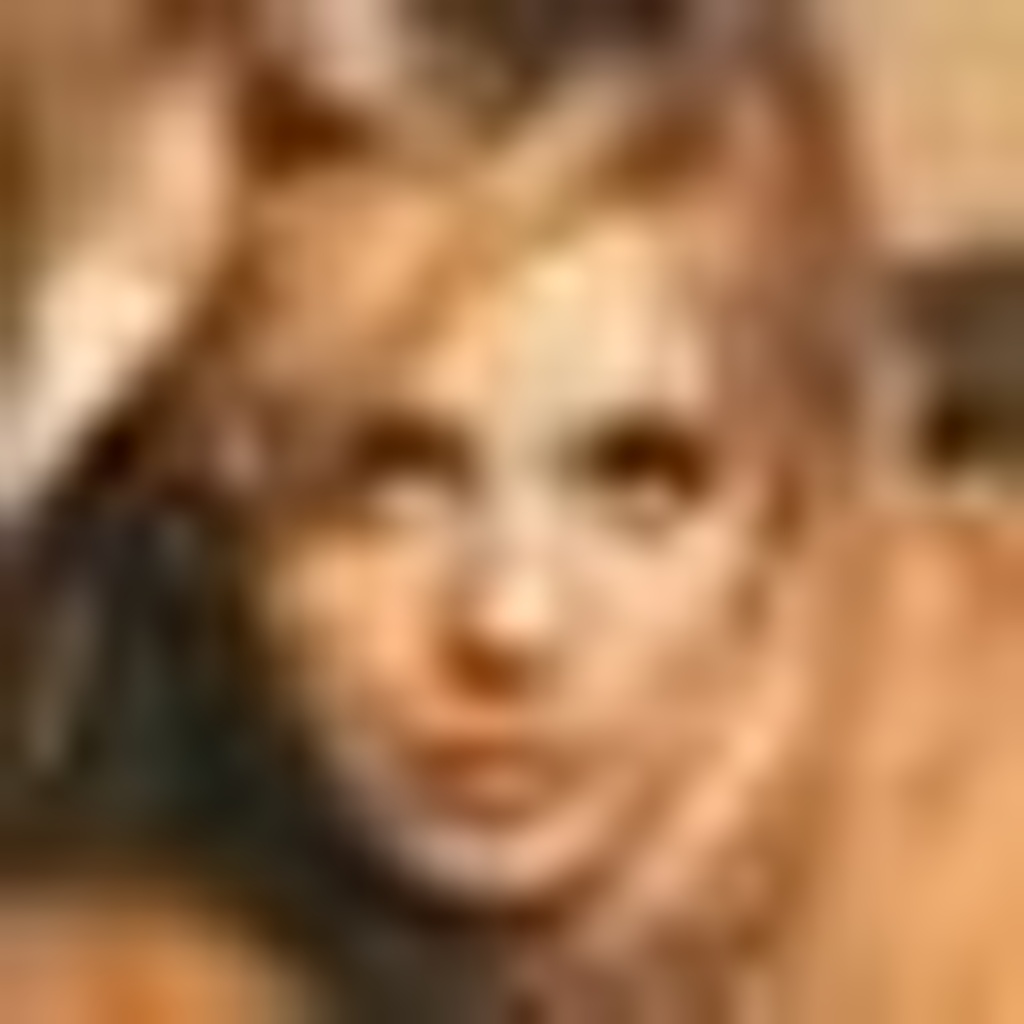}\vspace{4pt}
\includegraphics[width=1\linewidth]{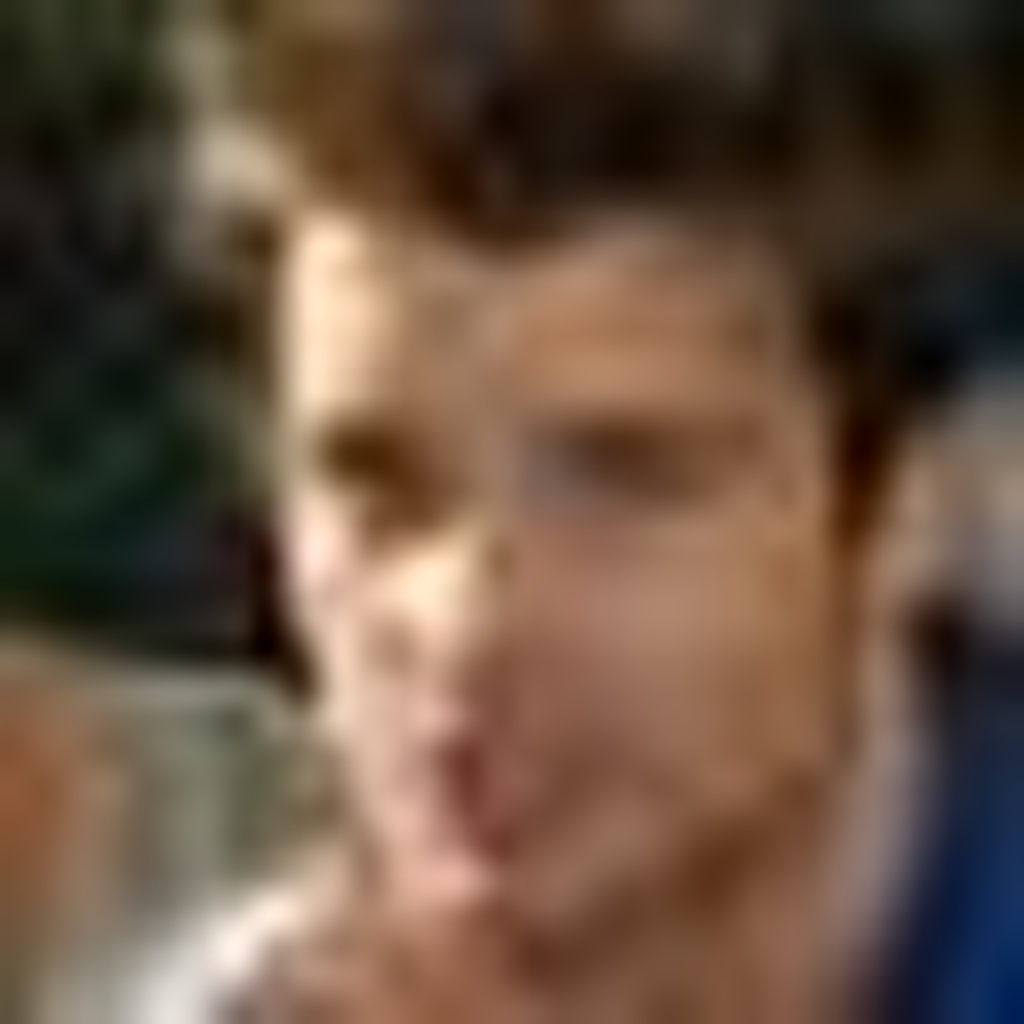}\vspace{4pt}
\end{minipage}}
\subfigure[PULSE\cite{2020PULSE}]{
\begin{minipage}[b]{0.2\linewidth}
\includegraphics[width=1\linewidth]{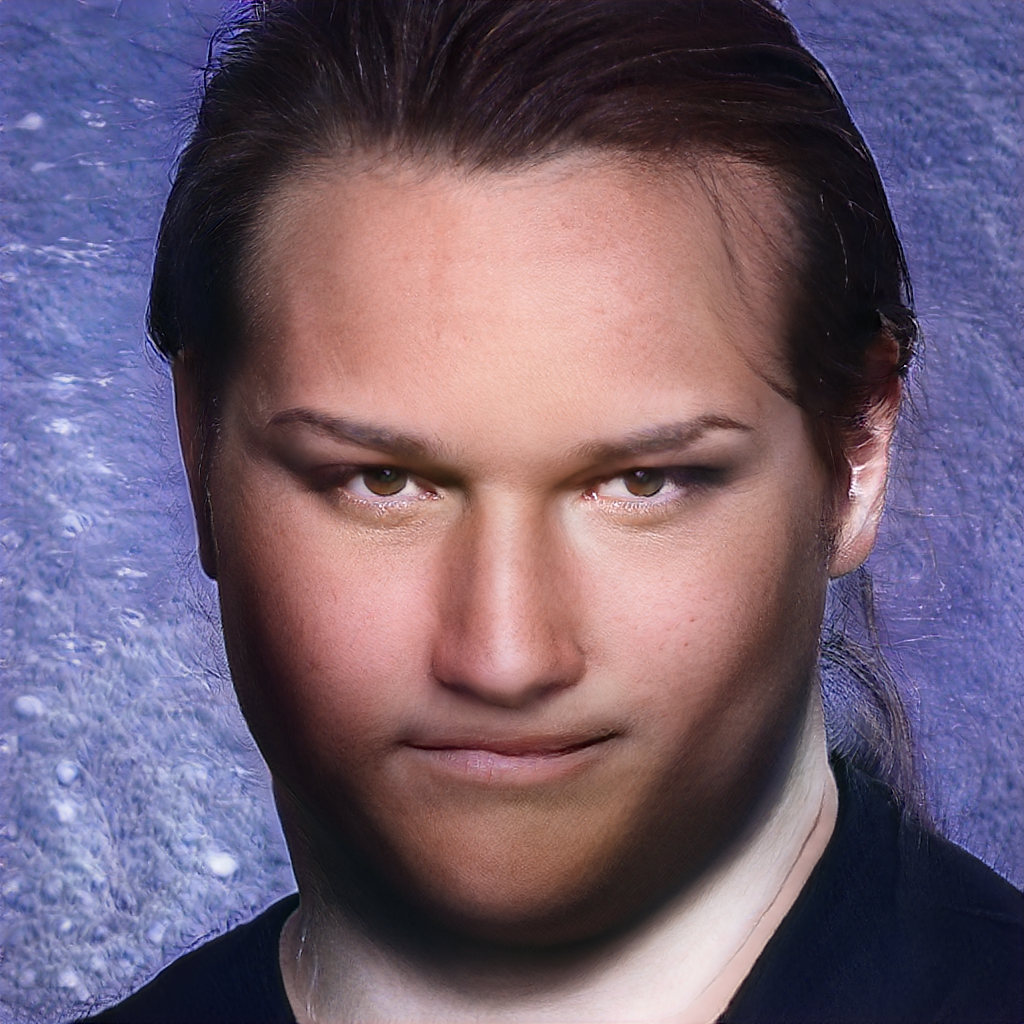}\vspace{4pt}
\includegraphics[width=1\linewidth]{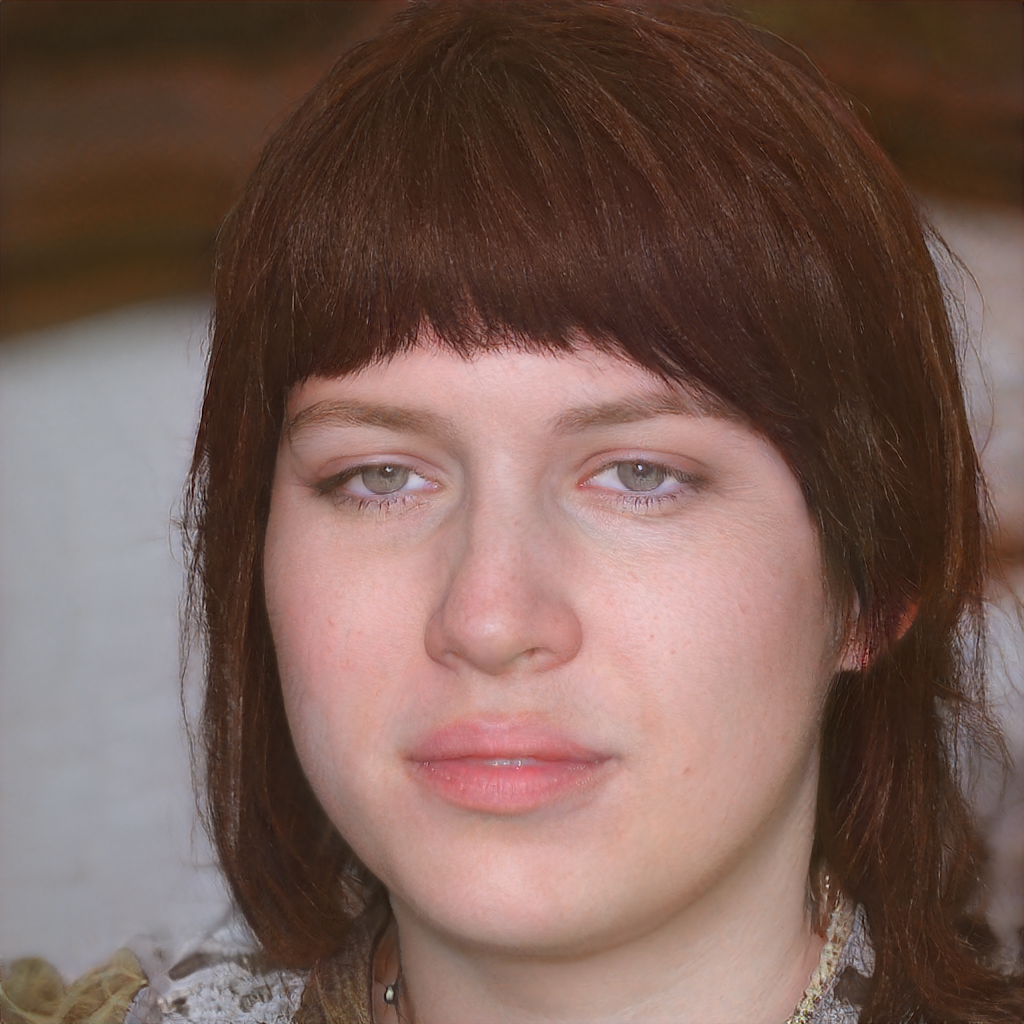}\vspace{4pt}
\includegraphics[width=1\linewidth]{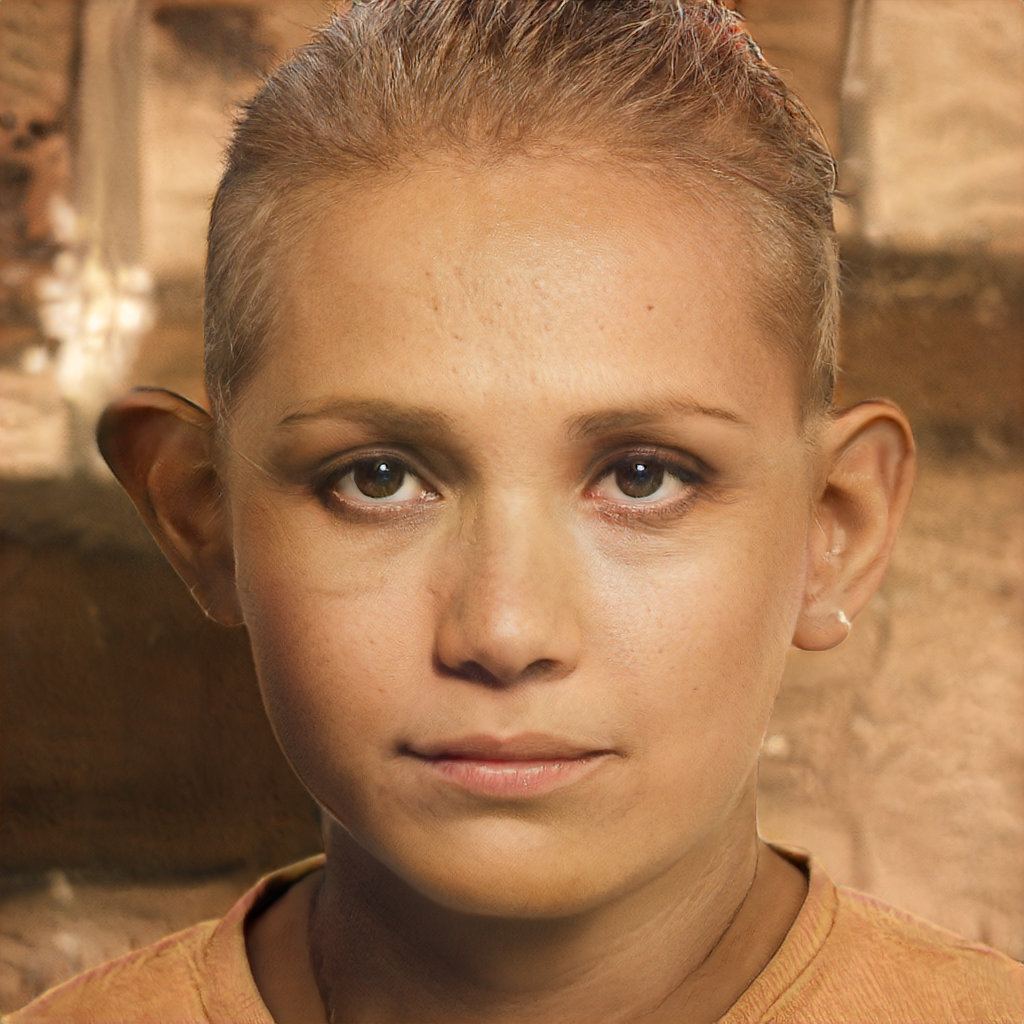}\vspace{4pt}
\includegraphics[width=1\linewidth]{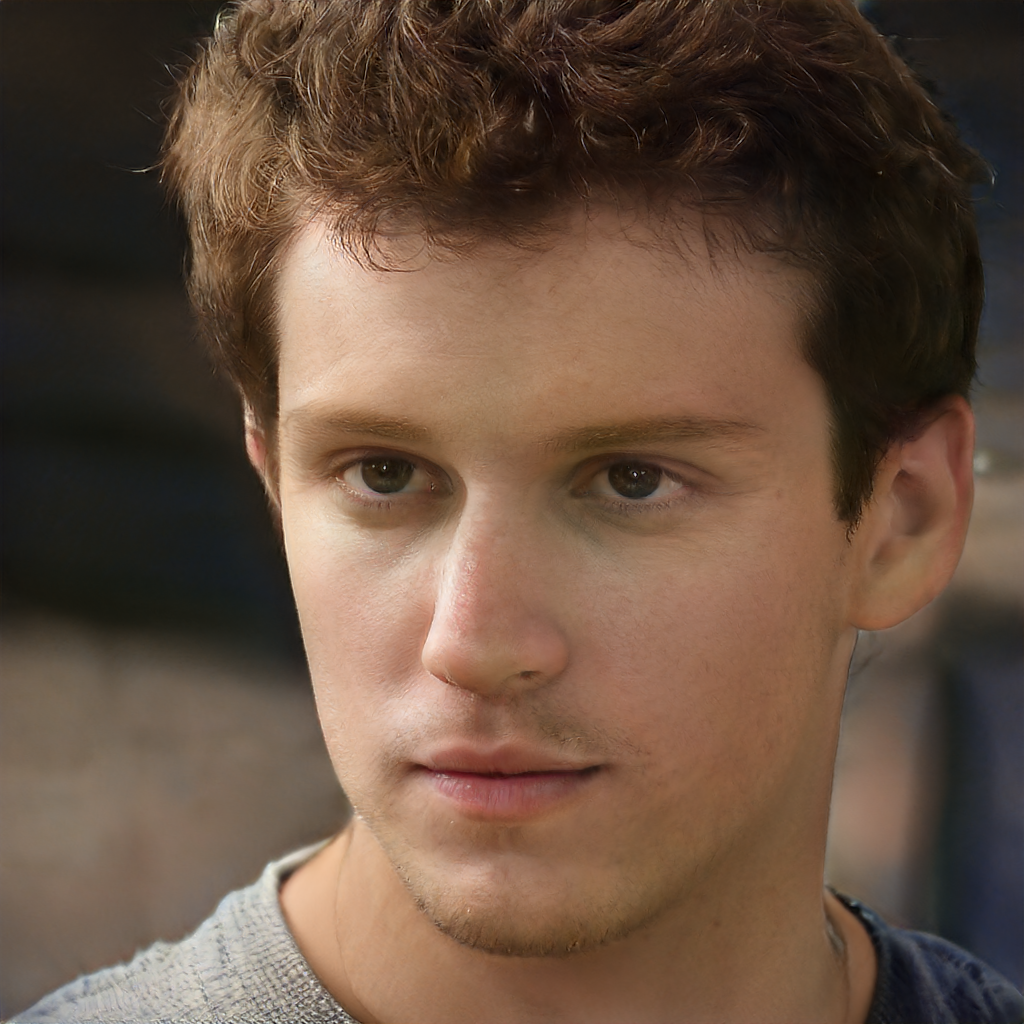}\vspace{4pt}
\end{minipage}}
\subfigure[Ours]{
\begin{minipage}[b]{0.2\linewidth}
\includegraphics[width=1\linewidth]{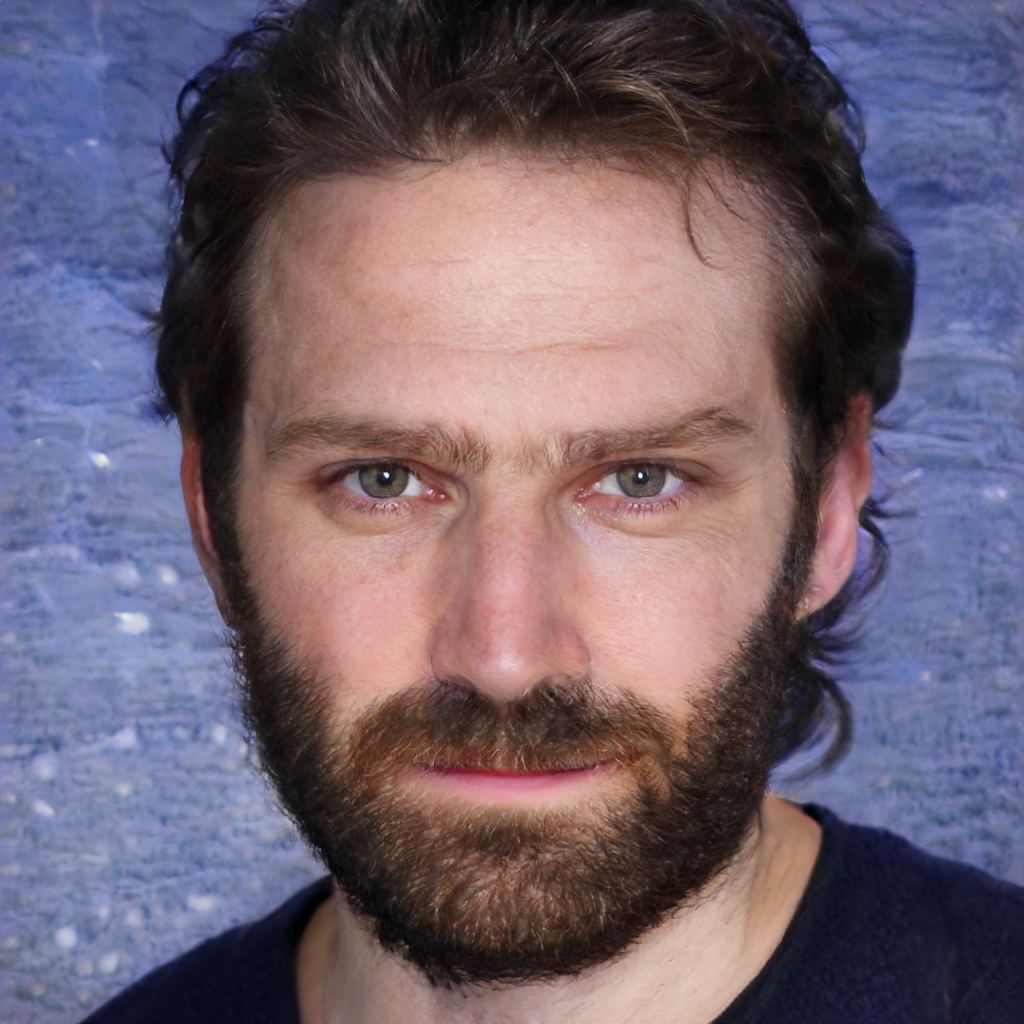}\vspace{4pt}
\includegraphics[width=1\linewidth]{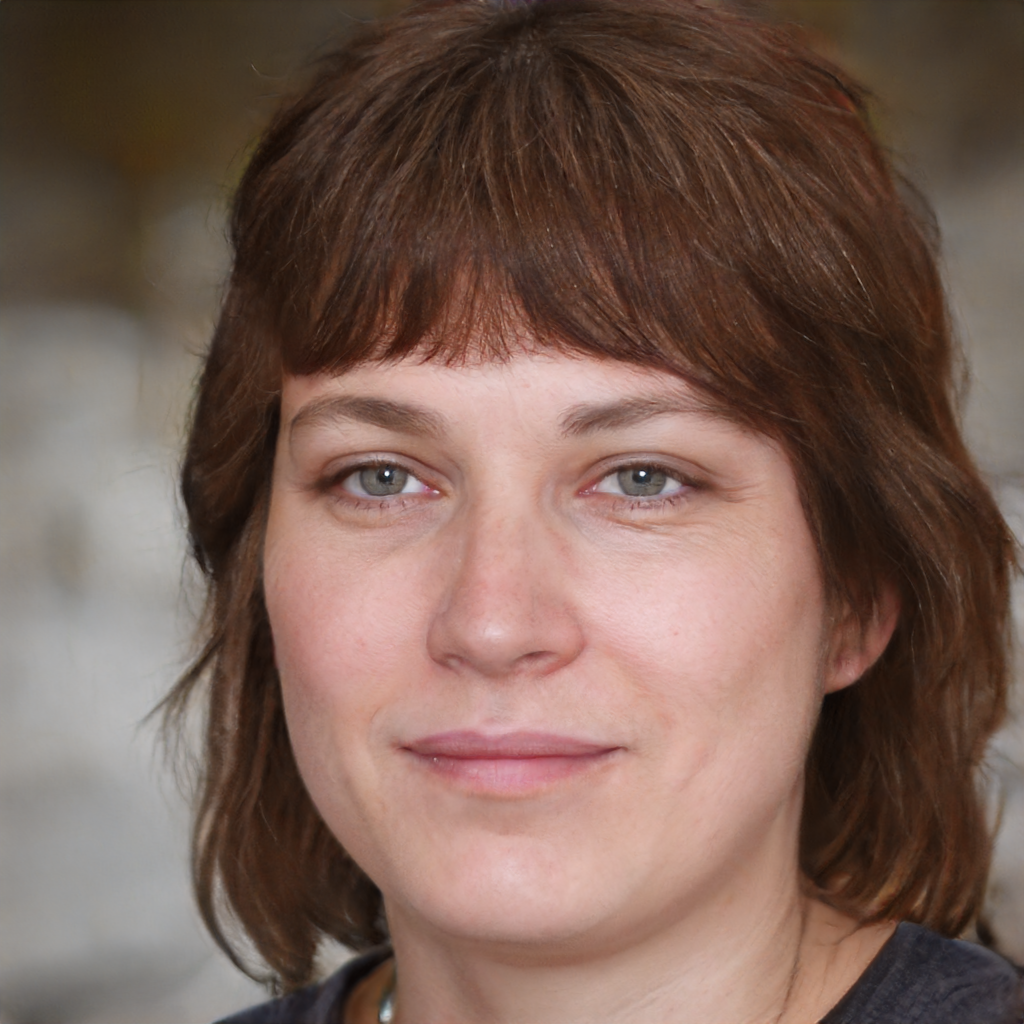}\vspace{4pt}
\includegraphics[width=1\linewidth]{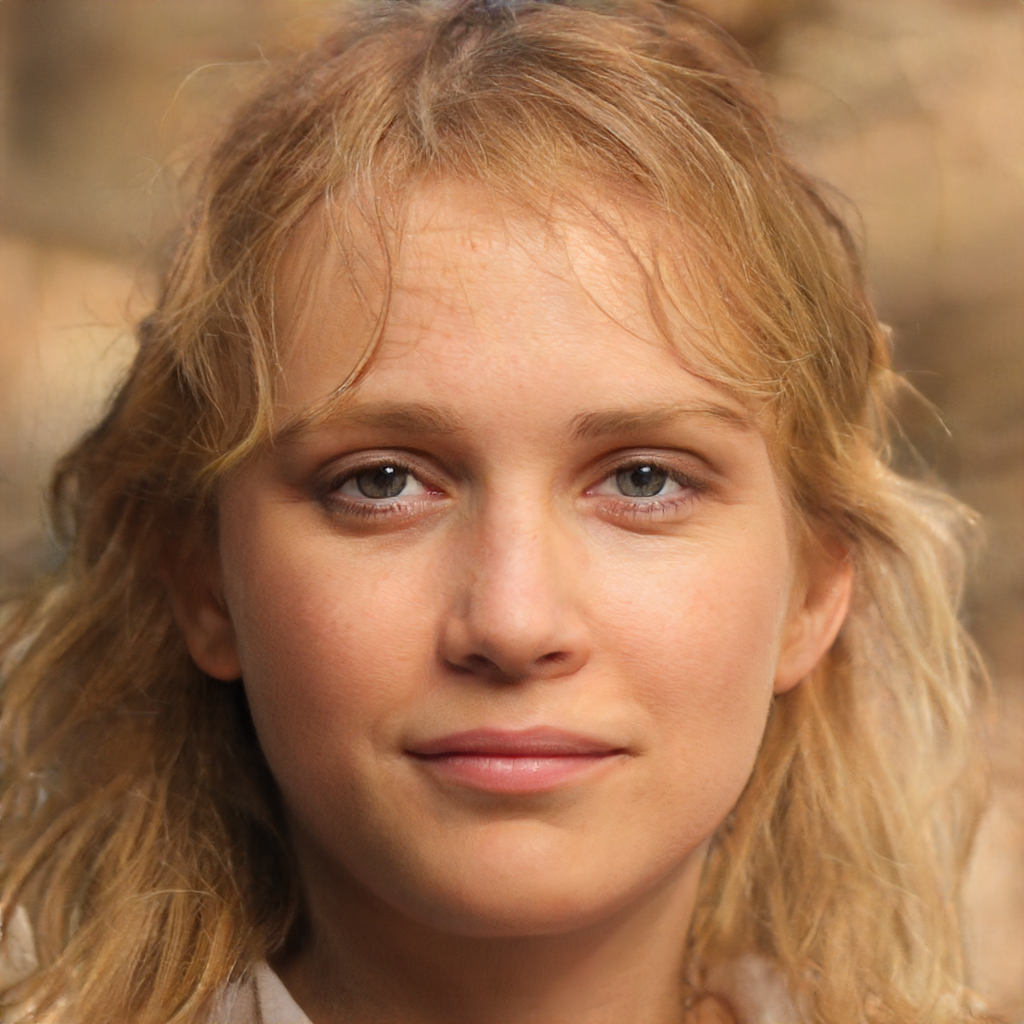}\vspace{4pt}
\includegraphics[width=1\linewidth]{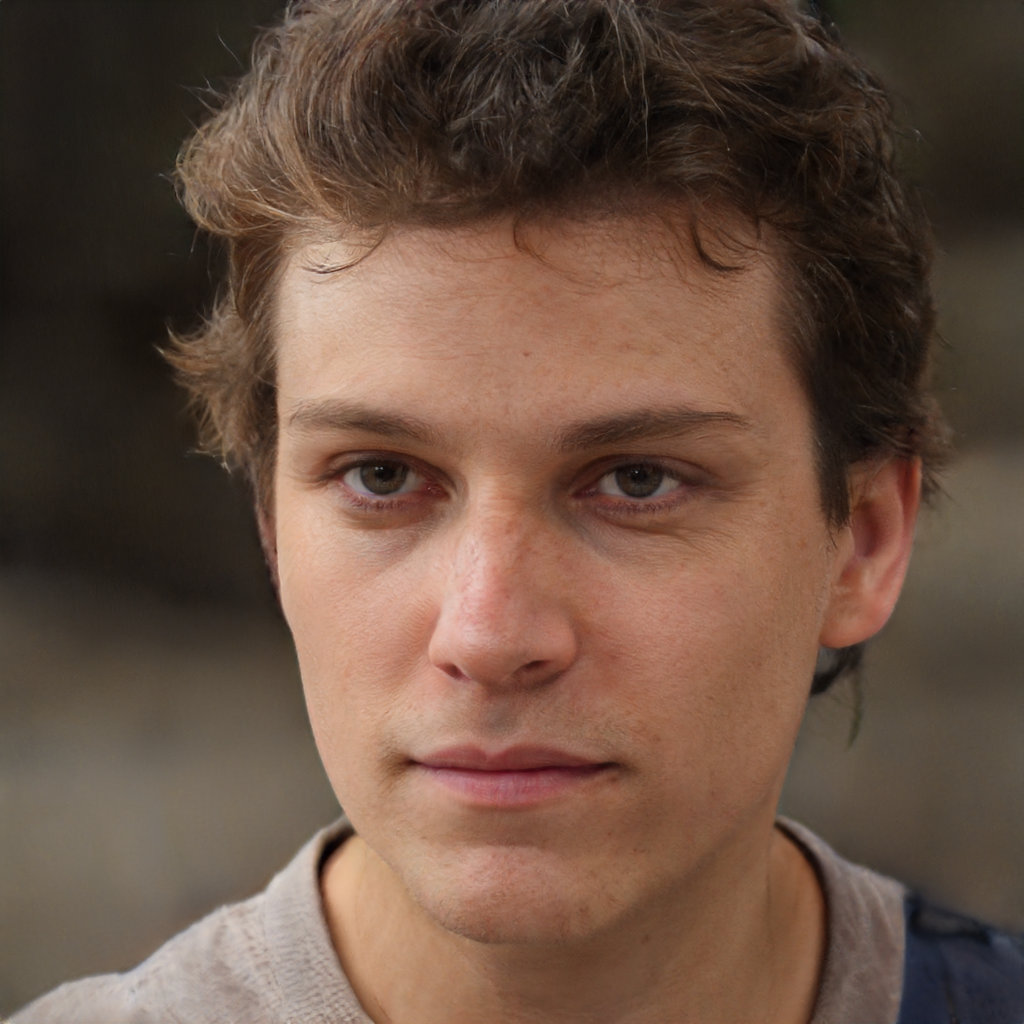}\vspace{4pt}
\end{minipage}}

\caption{Comparison of restored images optimized with/without the MMD loss. The incorporation of the MMD loss outputs more natural images.}
\label{super}
\end{figure}

\begin{figure}
\centering
\includegraphics[width=.9\linewidth]{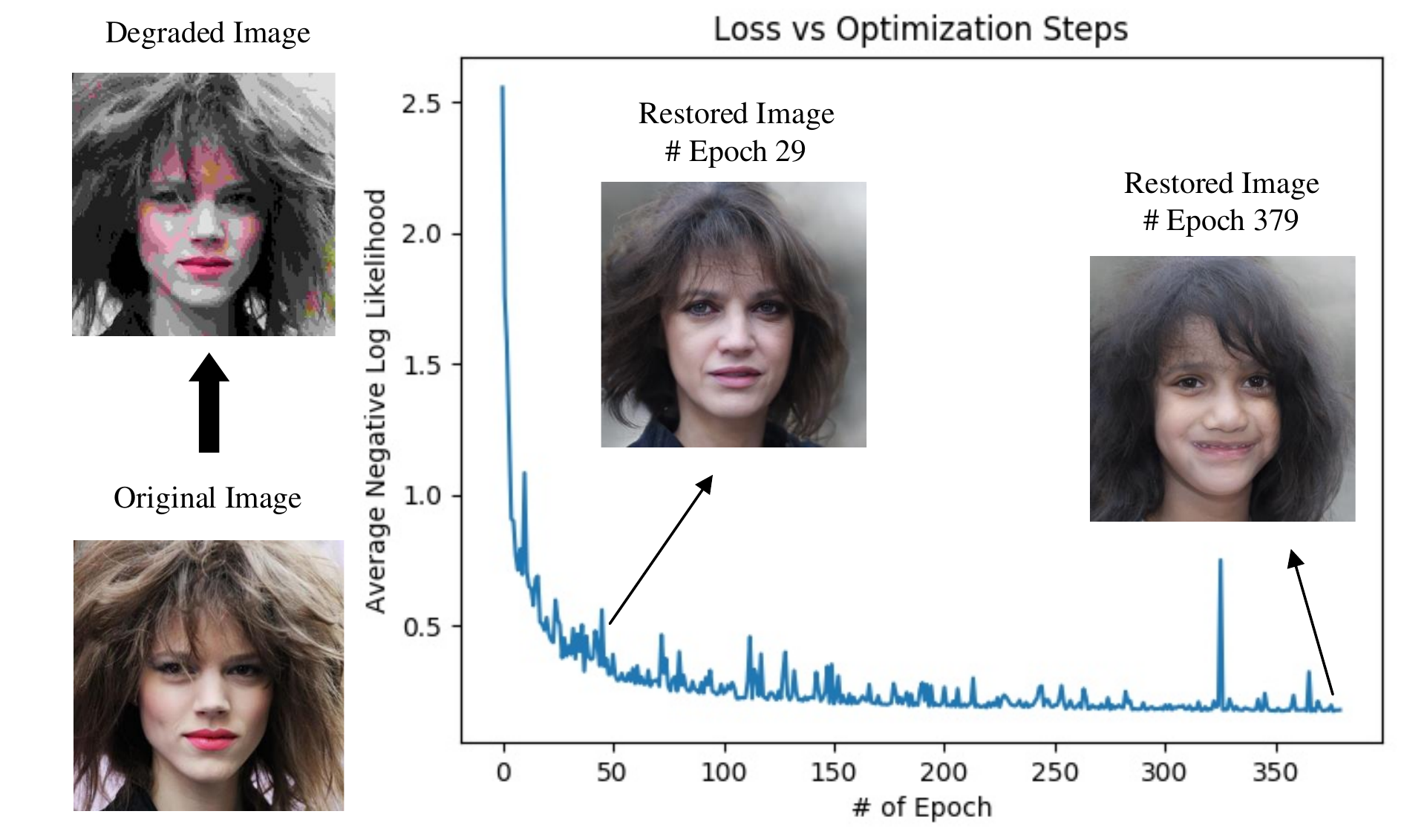}\vspace{4pt}
\caption{Loss curve when training the degradation model. The NLL is averaged on test set. We also provide the restored example at different epochs.}
\label{loss}
\end{figure}

\begin{table}
\begin{center}
\begin{tabular}{|c|c|c|} 
\hline
 & paired& unpaired\\
\hline
 NLL(bits/dim) & 0.39 & 2.26\\
\hline
\end{tabular}
\end{center}
\caption{ The average NLL score for paired and unpaired images. "paired" means that the input image is dagraded from the conditional image. "unpaired" means that the input image is dagraded from another image and hence has different content with the conditional image. }
\label{table1}
\end{table}

\begin{figure}
\centering
\subfigure{
\begin{minipage}[b]{0.139\linewidth}
\includegraphics[width=1\linewidth]{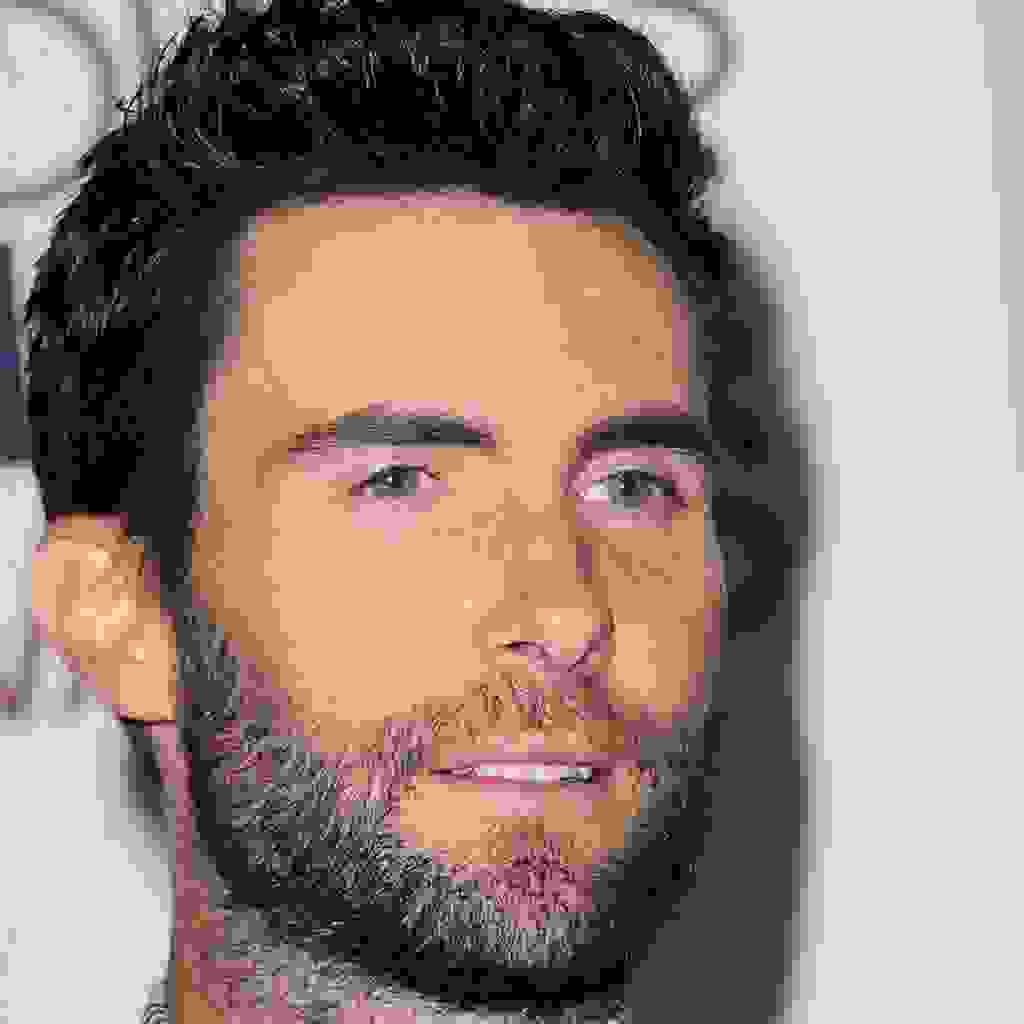}\vspace{4pt}
\centerline{(a) Reference}
\end{minipage}}
\subfigure{
\begin{minipage}[b]{0.139\linewidth}
\includegraphics[width=1\linewidth]{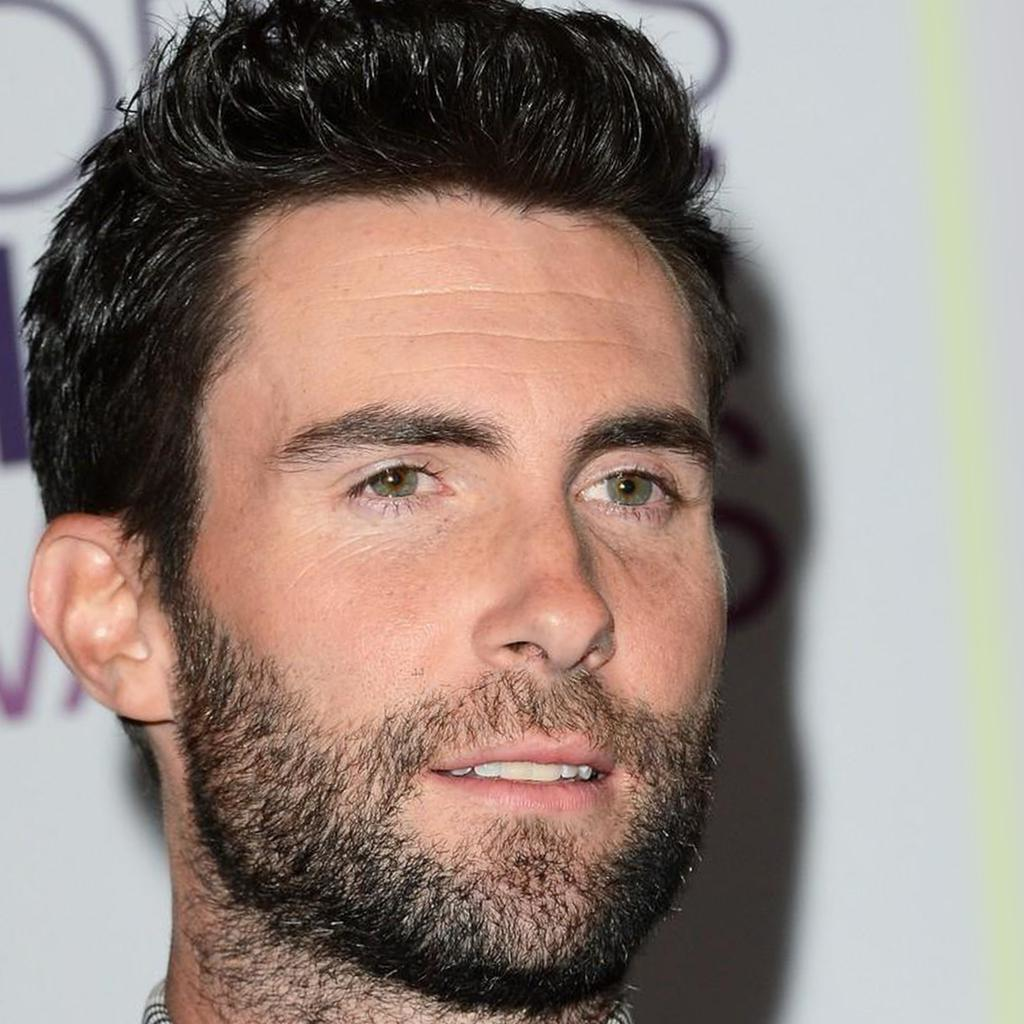}\vspace{4pt}
\centerline{(b) paired: 0.41}
\end{minipage}}
\subfigure{
\begin{minipage}[b]{0.139\linewidth}
\includegraphics[width=1\linewidth]{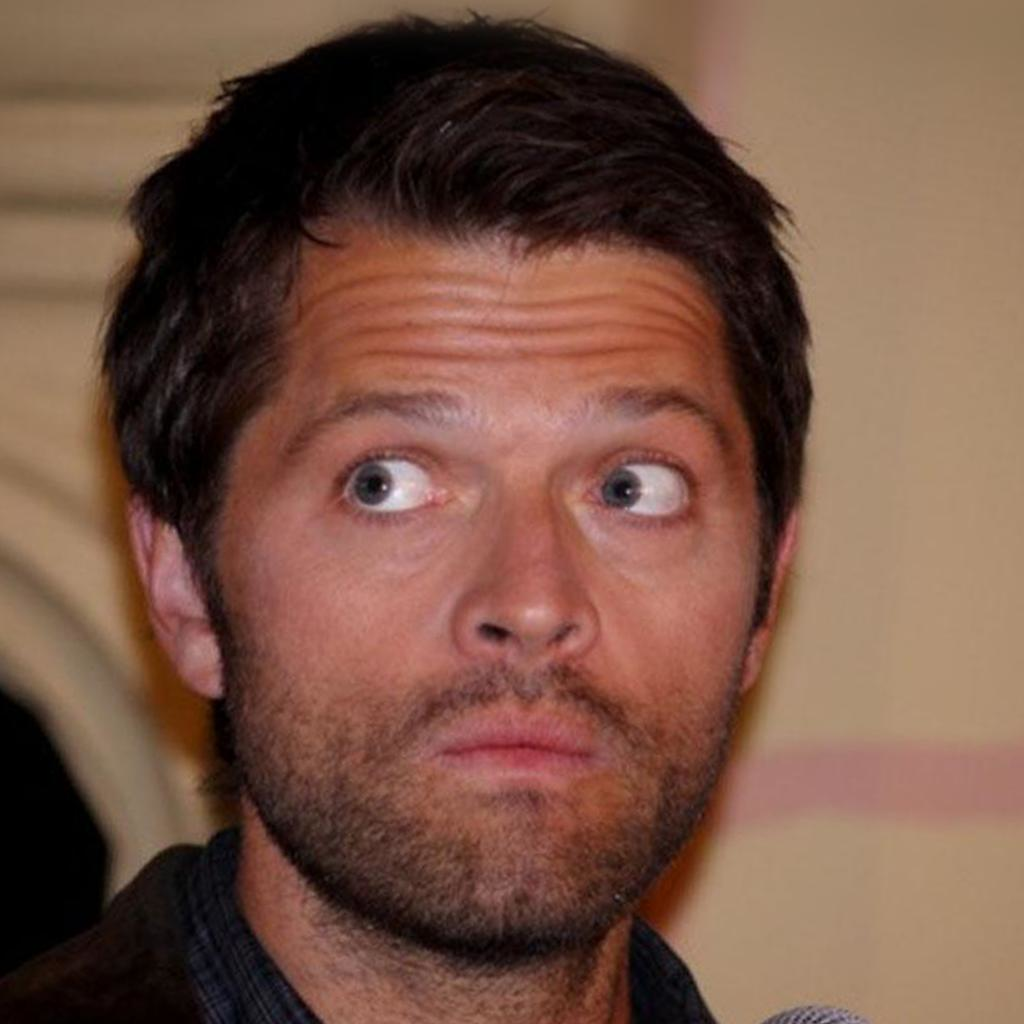}\vspace{4pt}
\centerline{(c) unpaired: 2.48}
\end{minipage}}
\subfigure{
\begin{minipage}[b]{0.139\linewidth}
\includegraphics[width=1\linewidth]{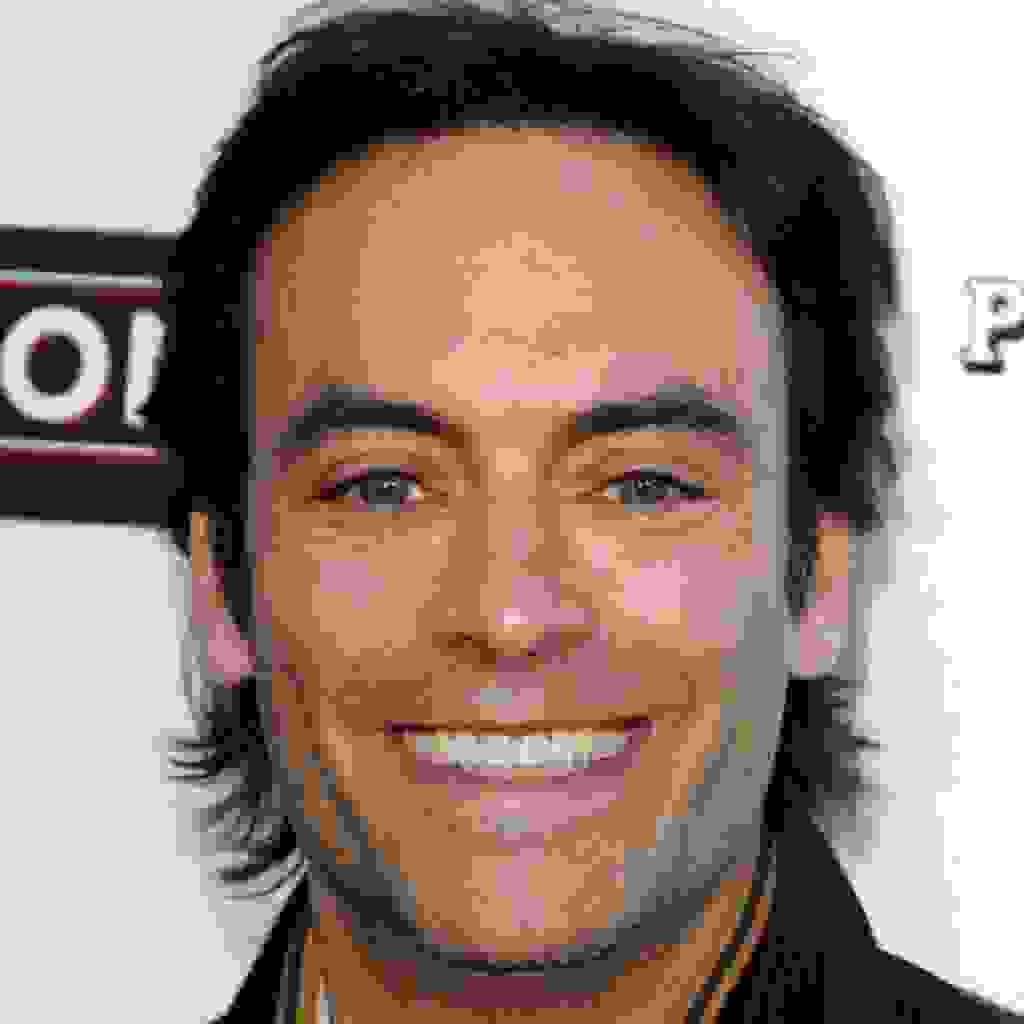}\vspace{4pt}
\centerline{(d) Reference}
\end{minipage}}
\subfigure{
\begin{minipage}[b]{0.139\linewidth}
\includegraphics[width=1\linewidth]{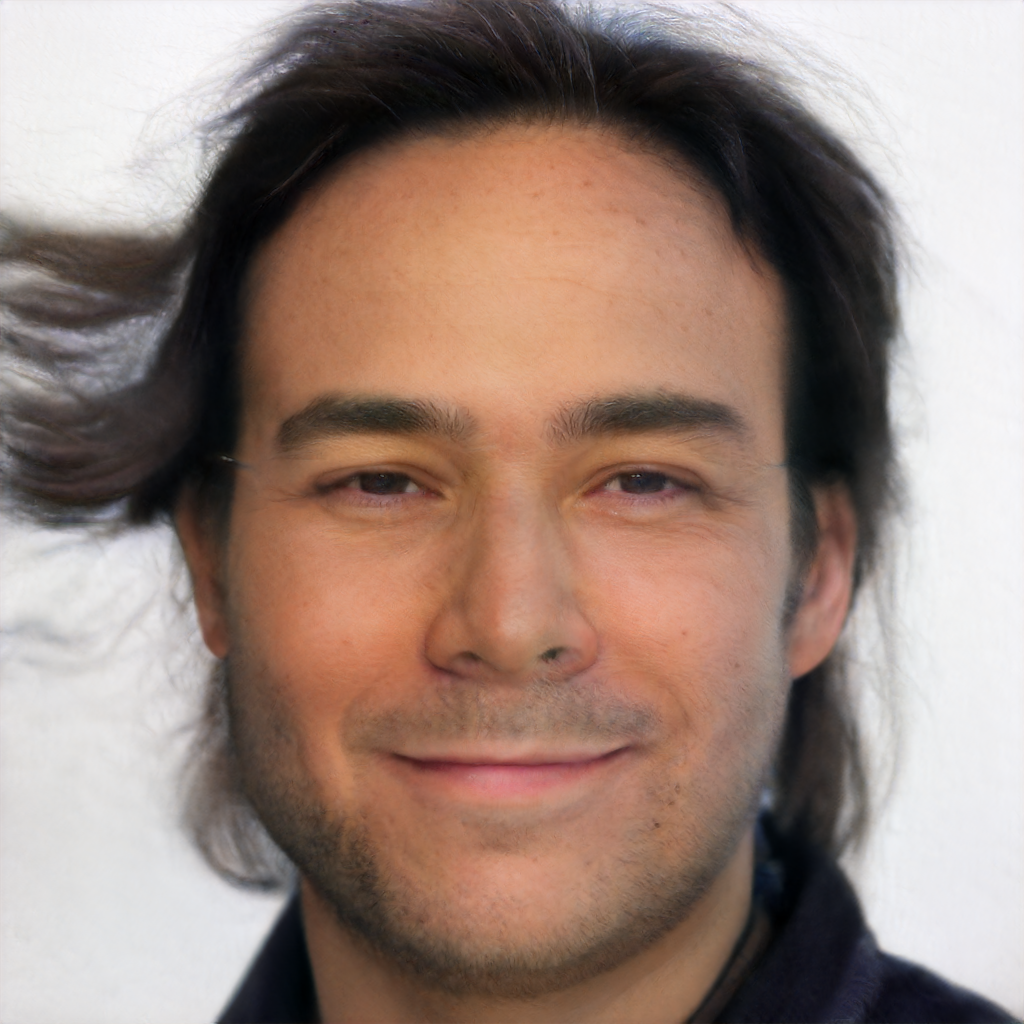}\vspace{4pt}
\centerline{(e) wo MMD: 0.69}
\end{minipage}}
\subfigure{
\begin{minipage}[b]{0.139\linewidth}
\includegraphics[width=1\linewidth]{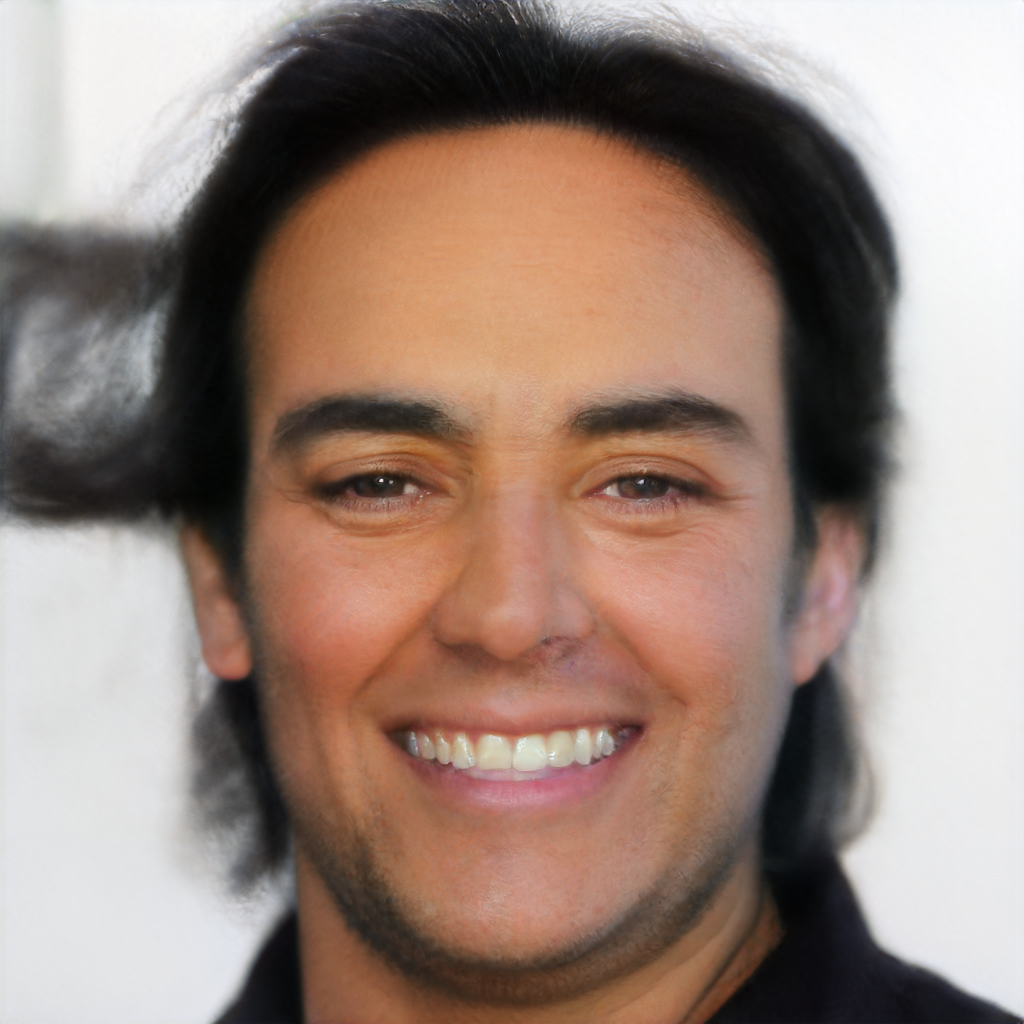}\vspace{4pt}
\centerline{(f) w MMD: 0.61}
\end{minipage}}
\caption{Left: The NLL for paired and unpaired images. Lower NLL suggests  higher probability of the (high,degraded) pair. Left: (a) is the degraded image from (b), while (c) is another high-quality image.  Right: (d) is a degraded image, while (e) is the restored image using PULSE and (f) is restored with MMD loss. The degradation model suggests (f) as a closer solution to (d). This is consistent with the subjective judgment. Furthermore, this demonstrates that the MMD loss leads to a better optimization. }
\label{prob}
\end{figure}

\subsection{Image Super-resolution}

To demonstrate the effectiveness of the MMD loss, we firstly conduct experiments on image super-resolution. We adopt the same method used in \cite{2020PULSE} for this task. The only difference is that we replace the StyleGAN with the improved version  for a better baseline model. The examples are shown in Figure \ref{super}. The low-resolution images are with size $32\times32$ while the high-resolution is $1024\times1024$.
The results of PULSE is optimized with the spherical gradient descent \cite{2020PULSE}, while our method incorporates the additional MMD term. 
All other settings are the same. We can see that even with a better StyleGAN model, PULSE still produces unnatural images. The failures make it unreliable in practical systems.
The failure could either origin from the divergence with prior distribution or the unconsistency between styles (they should be identical in original models).
Whatever the reason, our model successfully fixes it. With the MMD loss restricting the latent embedding, we are able to generate more natural and high-quality images.

\subsection{Degradation Estimation}
We train the degradation model in a split of the CelebA HQ dataset \cite{celeba-hq}, which contains 20,000 randomly selected images. The rest is referred  as test set. The corresponding degraded images are generated with the JPEG \cite{JPEG} standard. For memory issues, we resize both the conditional and degraded images  to  $256\times256$.
  
\noindent \textbf{Effectiveness}.
Figure \ref{loss} shows the loss curve in terms of average negative log likelihood (NLL) computed on test set. In existence of conditional image, the model achieves 0.5 bits/dim after about 20 epochs. 
To demonstrate the effective of the degradation model on distinguishing the matched images from unmatched, we compute the average NLL score on paired and unpaired images and report corresponding results in Table \ref{table1}. As shown, the model achieves 0.39 bits/dim on paired (high,degraded) images and 2.26 bits/dim on unmatched pairs. An example is given in Figure \ref{prob}. The degradation model can to some extent estimate the similarity between degraded image $I_L$ and conditional image $I_H$, and give the explicit probability of $P_{I_L|I_H}$.
To further test the model ability in distinguishing small differences, we run the baseline algorithm with/without the MMD loss to produce different high-quality images similar to the degraded image. And then compute the NLL for corresponding pairs. The result is shown in Figure \ref{prob}. We can easily tell that Figure \ref{prob}(f) is more likely the original image than Figure \ref{prob}(e) due to the same "smile" attribute with the reference image. The degradation model also prefers Figure \ref{prob}(f) through a smaller NLL score. 
We attribute part of the ability in detecting small differences to the local property of the degradation model. The local design enables a bottom to top fashion and models the overall probability as the aggregation of local similarities.

\begin{figure}
\centering
\includegraphics[width=1\linewidth]{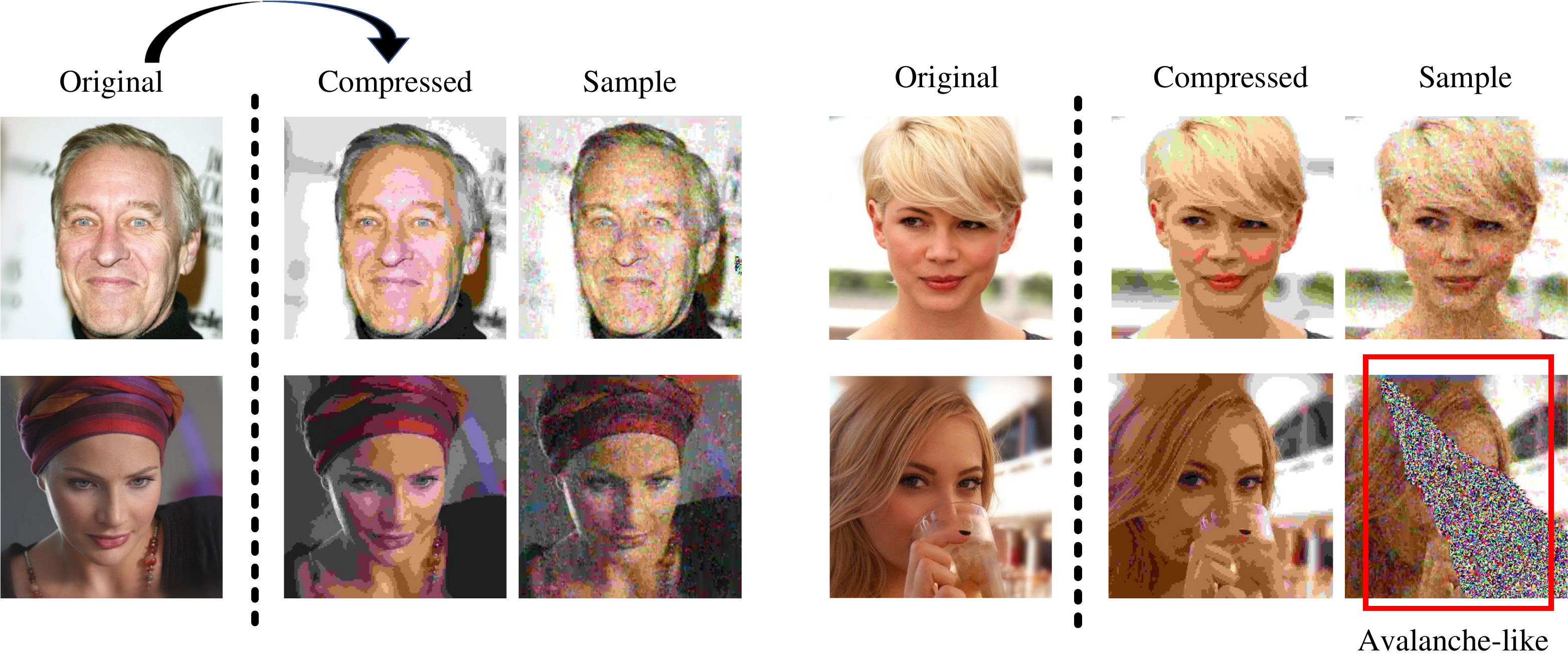}\vspace{4pt}
\caption{Samples drawn from the degradation model. The avalanche-like areas in the last sample illustrates that pixel by pixel sampling could lead to a chain reaction.}
\label{sample}

\end{figure}

\noindent \textbf{Sampling from Degradation Model}. An explicit modeling of the degradation distribution allows for sampling from corresponding process. We give several samples drawn from the trained model in Figure \ref{sample}. The pixel by pixel sampling takes hours to sample single image and is only used for visualization. Compared to the compressed images, we find that the sampled images do exhibit similar patterns, like the degraded color in the faces. Moreover, the original structure and textures are well preserved. They are both the desired properties.
However, the samples lack kind of spatial consistence and look noisy. On one hand, the pixel by pixel sampling lack explicit constraint on continuity.
On the other hand,  this is the direct result of the local design. 
Another consequence of the pixel by pixel sampling is the avalanche-like area shown in the last sample in Figure \ref{sample}. A sampling on pixels with very low probability would leads to a chain reaction, from a point to half of the image. We find sampling from degradation model gives a very clear illustration of such mechanism, which is not available in previous work \cite{pixelcnn,pixelcnn++}.  
 Such  phenomenon is not troublesome here since our model is not intended for generating images.  

\noindent \textbf{The NLL issue}.  During the experiments, we find that a lower NLL score does not necessarily imply a better degradation model. We test the model trained on 29 epoch and 379 epoch, with NLL 0.45 bits/dim and 0.18 bits/dim respectively. We find the restored images looking strange with the latter model, as shown in Figure \ref{loss}. Moreover, adding the degradation model does not bring a lower NLL. 
We attribute this to two reasons. Firstly, previous work has realized that the NLL is not consistent with the visual quality \cite{pixelcnn++}.
Secondly, as noted before, the autoregressive model suffers from severe overfitting. 
Therefore, a overtrained model may not generalize well, especially for a conditional version.

\begin{table}
\begin{center}
\scriptsize
\begin{tabular}{|c|c|c|c|c|c|c|c|c|c|c|c|} 
\hline
\multirow{2}{*}{Name} & \multicolumn{7}{c|}{Settings }& \multicolumn{2}{c|}{Score} & \multicolumn{2}{c|}{Score (Resize)} \\
\cline{2-12}
 &StyleGAN V& mean Init& MMD & Spherical & pixelcnn coff & Epoch & step & mean & median & mean & median\\
\hline\hline
&v1  & N & N & Y & - & - & 100 & 33.21 & 33.12 &17.17 &17.25\\ 
\hline
&v2 & N & N & Y & - & - & 100 & 31.82 & 32.15 &17.87 &17.93\\ 
&v2 & Y & N & Y & - & - & 100 & 35.11 & 34.94 &21.33 &21.32\\ 
&v2 & N & Y & Y & - & - & 100 & 35.66 & 35.20 &20.23 &20.22\\ 
Ablation&v2 & Y & Y & Y & - & - & 100 & 30.17 & 30.34 &9.78 &9.74\\ 
&v2 & Y & Y & Y & - & - & 1000 & 31.55 & 31.54 &11.94 &11.91\\ 
Study&v2 & Y & Y & N & - & - & 100 & 32.85 & 32.67 &15.05 &15.10\\ 
&v2 & Y & Y & Y & 1 & 29 & 100 & 31.40 & 31.37 &10.28 &10.14\\
&v2 & Y & Y & Y & 1 & 379 & 100 & 30.41 & 30.64 &11.15 &11.11\\
&v2 & Y & Y & Y & 1 & 379 & 1000 & 31.79 & 31.65 &11.69&11.60\\
&v2 & Y & Y & Y & 10 & 29 & 100 & 30.57 & 30.69 &\textbf{9.37} &\textbf{9.23}\\
&v2 & Y & Y & Y & 10 & 379 & 100 & \textbf{29.41} & \textbf{29.97} &10.04 &10.05\\
&v2 & Y & Y & Y & 50 & 379 & 100 & 29.84 & 30.69 &10.56 &10.53\\
\hline
Original &- & - & - & - & - & - & - & 45.61 & 45.75 &13.67 &13.65\\
\hline
Compressed &- & - & - & - & - & - & - & 59.47 & 59.81 &54.57 &54.62\\
\hline
\end{tabular}
\end{center}
\caption{Mean and median score from RankIQA for various image sets (smaller is better). "Original" denotes images from CelebA-HQ dataset. "Compressed" denotes images after JPEG compression. "StyleGAN V": version of StyleGAN. "mean Init": Initialized from mean latent. "MMD": with MMD. "Spherical": Spherical gradient descent for latents. "pixelcnn coff": the cofficient for pixelcnn loss. "Epoch": the epoch number for training pixelcnn model. "step": optimization step.  "Score" and "Score (Resize)" are the two strategies used for computing the score.}
\label{Score}
\end{table}



\begin{figure*}
\centering
\subfigure[Original]{
\begin{minipage}[b]{0.16\linewidth}
\includegraphics[width=1\linewidth]{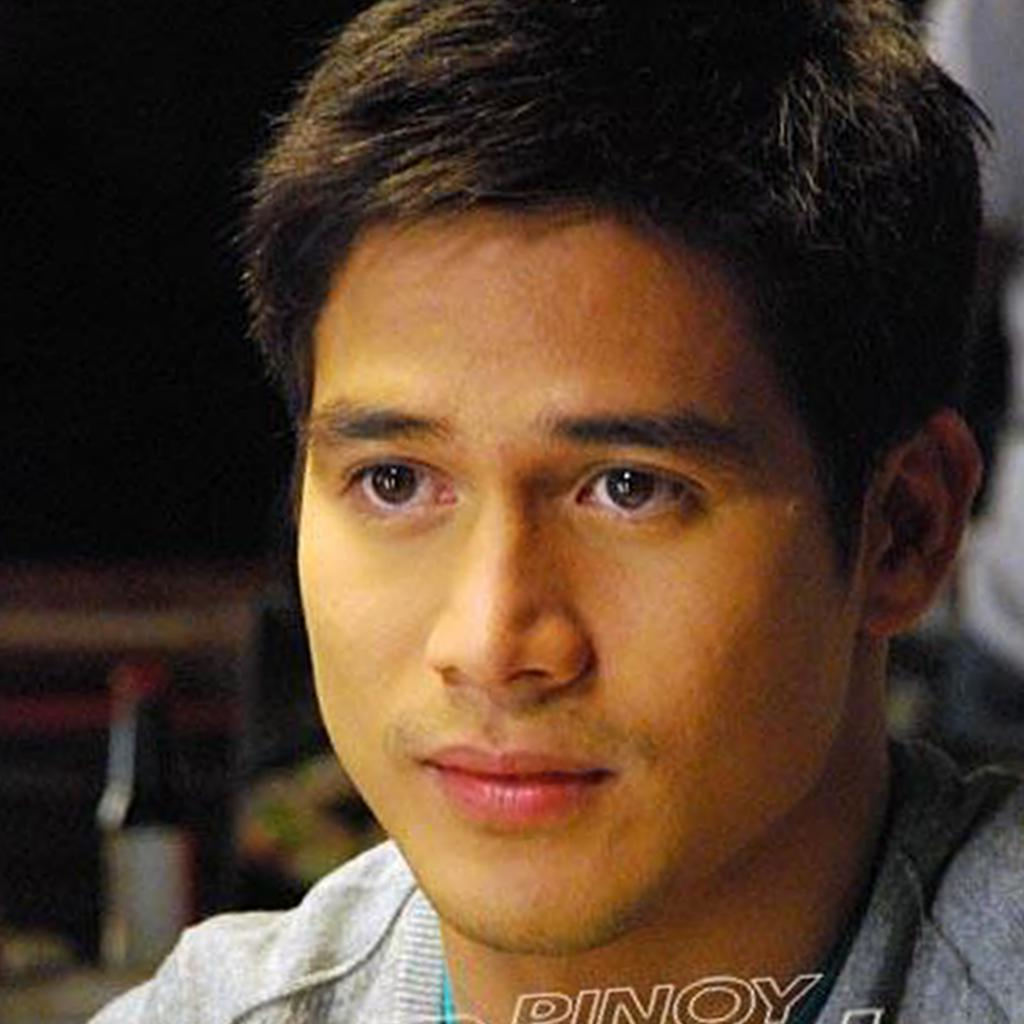}\vspace{4pt}
\includegraphics[width=1\linewidth]{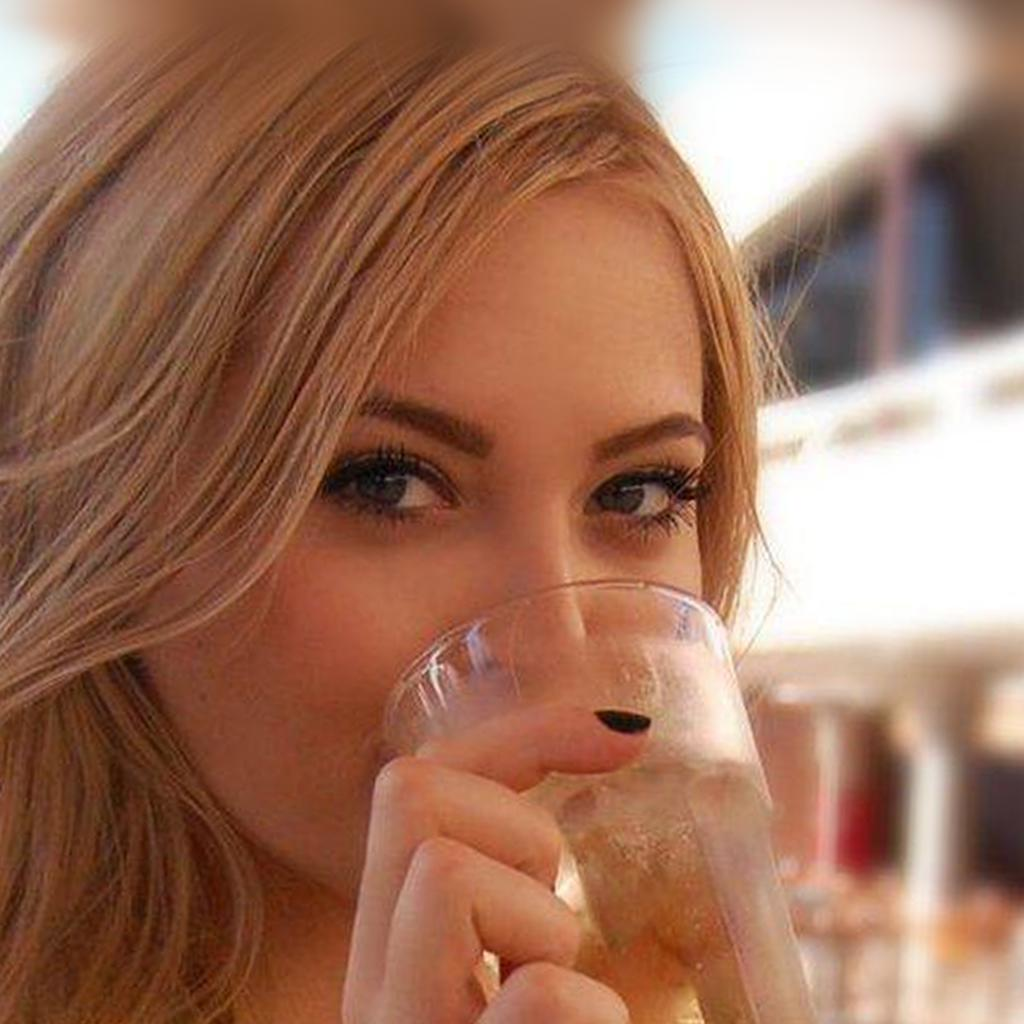}\vspace{4pt}
\includegraphics[width=1\linewidth]{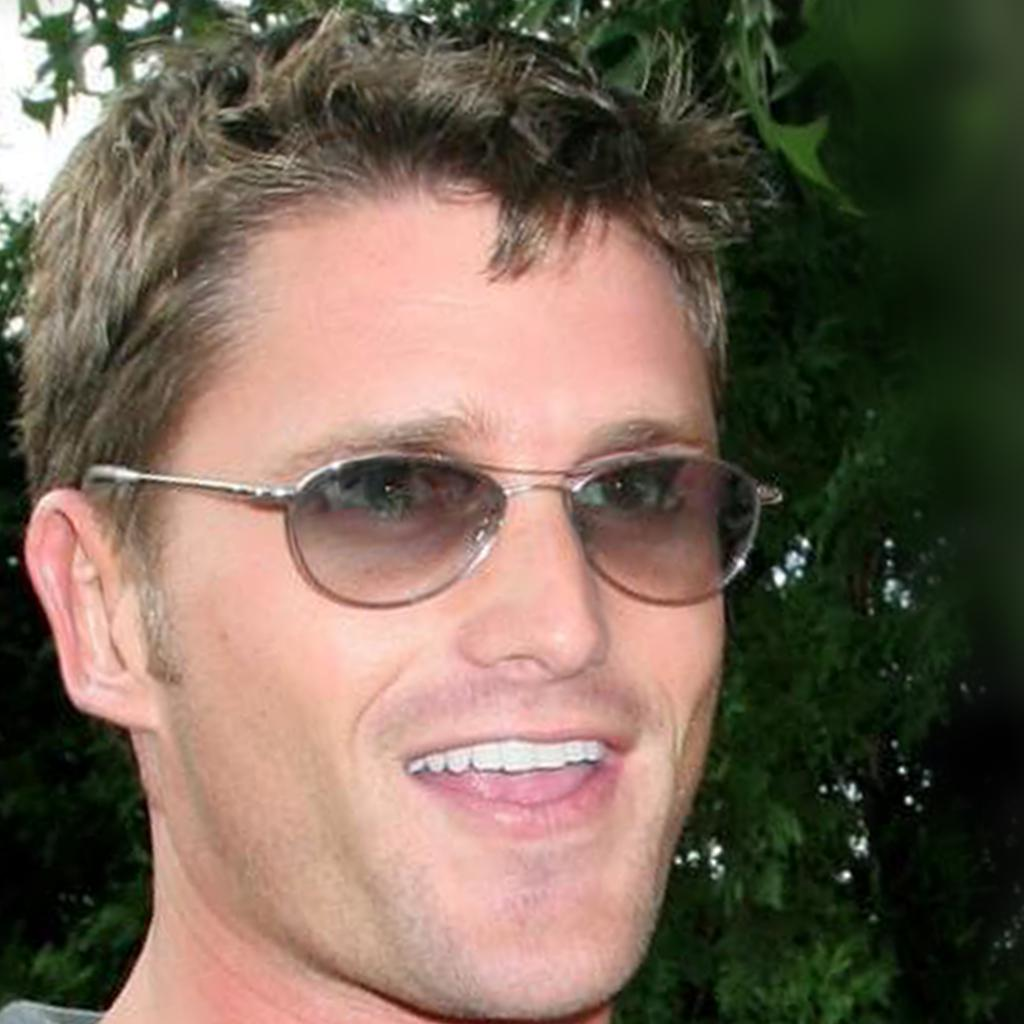}\vspace{4pt}
\includegraphics[width=1\linewidth]{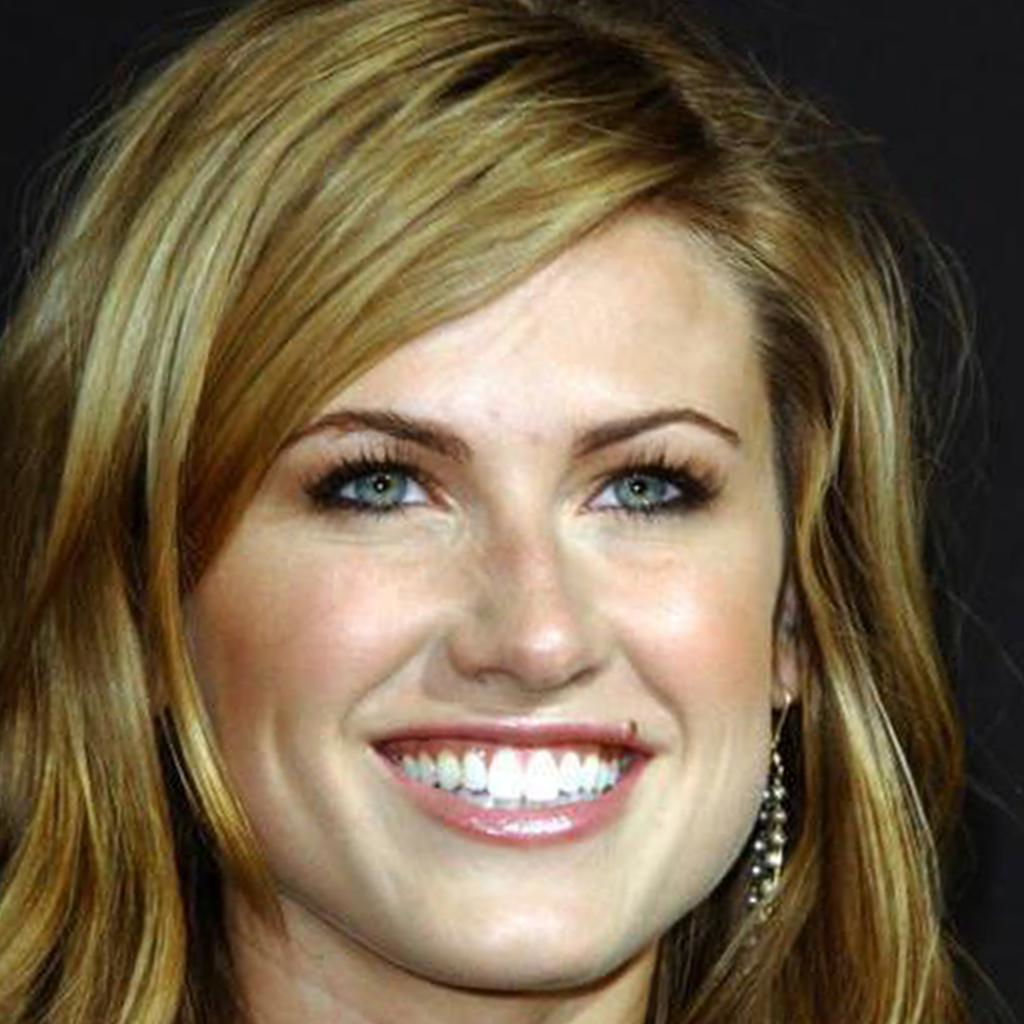}\vspace{4pt}
\includegraphics[width=1\linewidth]{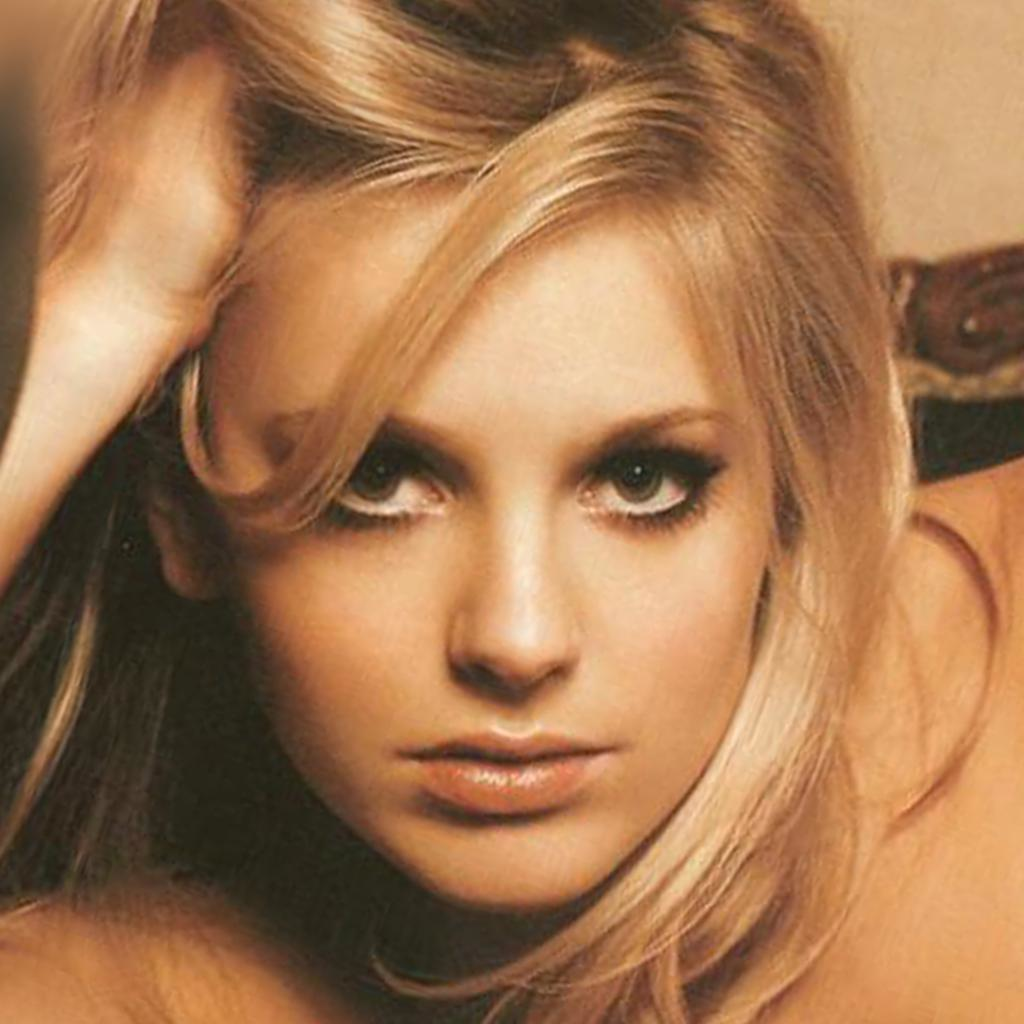}\vspace{4pt}
\includegraphics[width=1\linewidth]{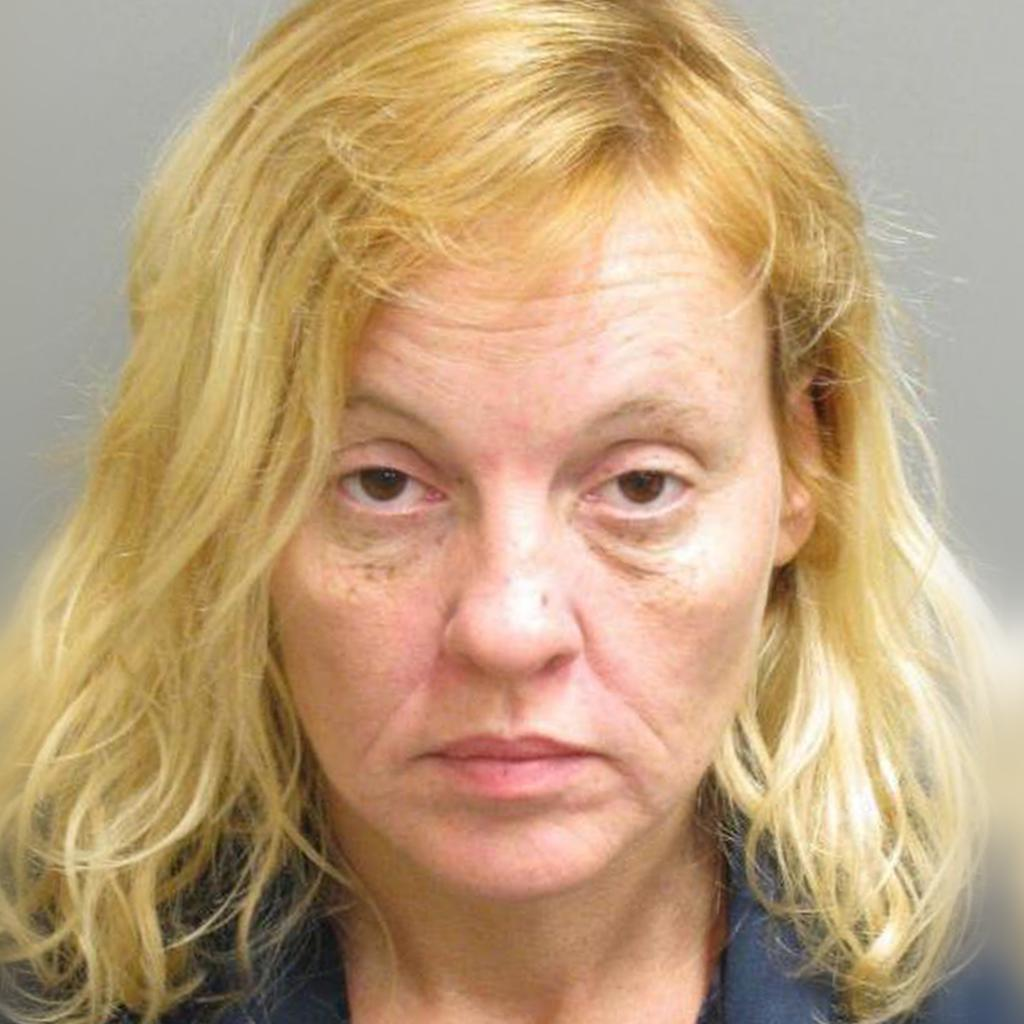}\vspace{4pt}
\end{minipage}}
\subfigure[Compressed]{
\begin{minipage}[b]{0.16\linewidth}
\includegraphics[width=1\linewidth]{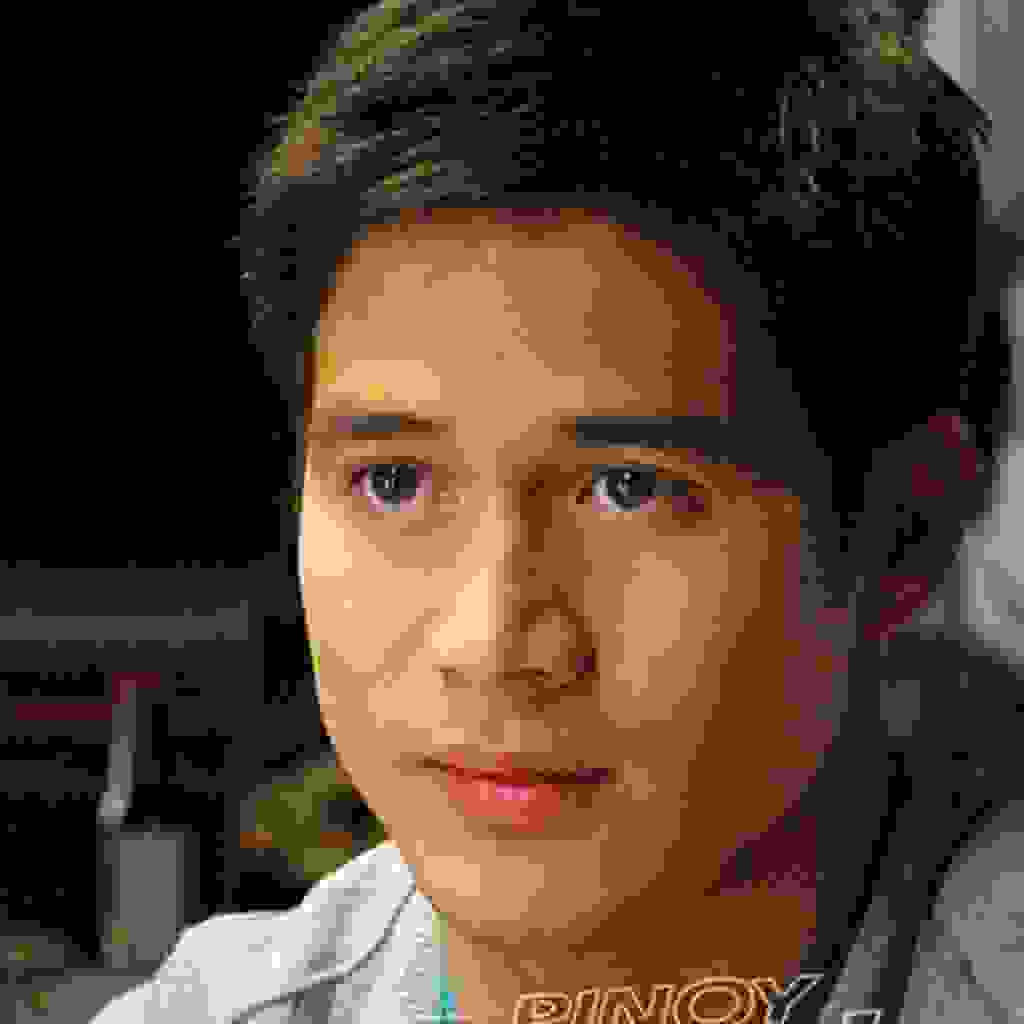}\vspace{4pt}
\includegraphics[width=1\linewidth]{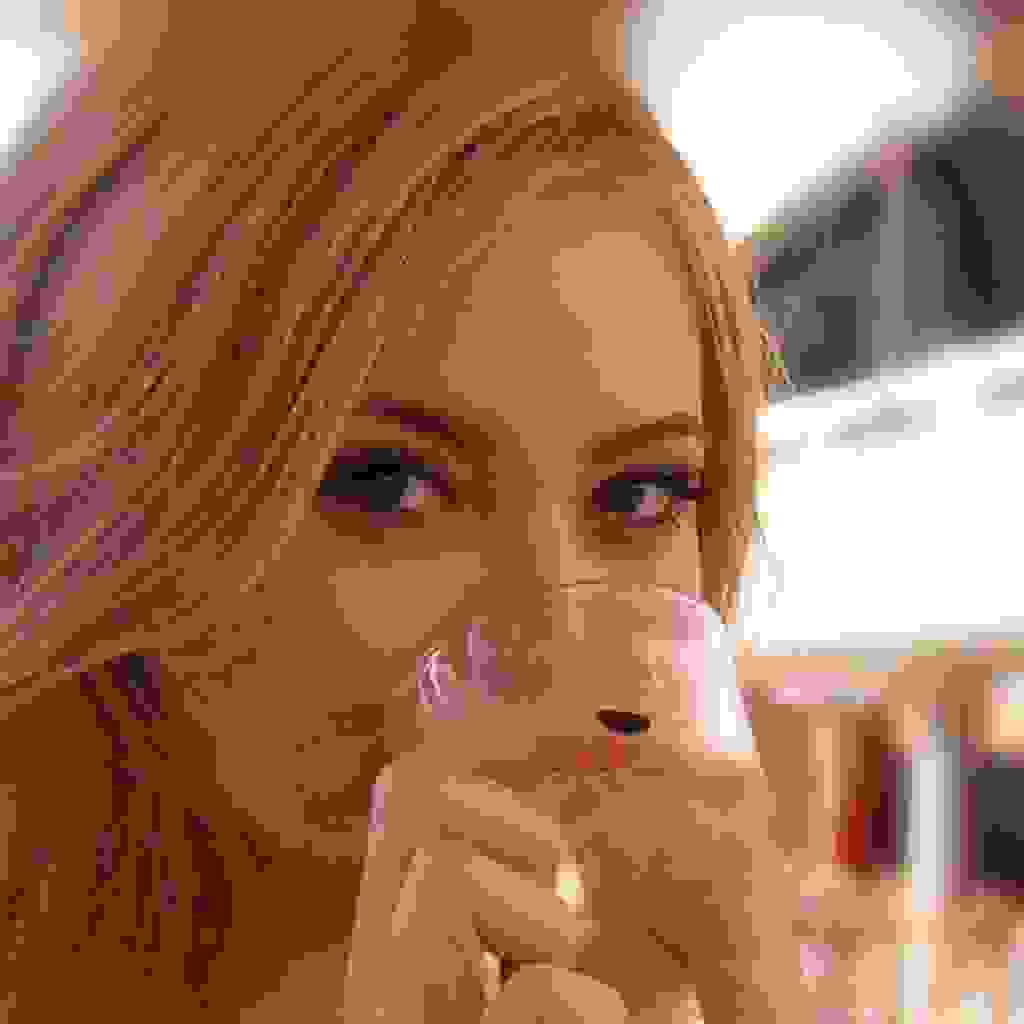}\vspace{4pt}
\includegraphics[width=1\linewidth]{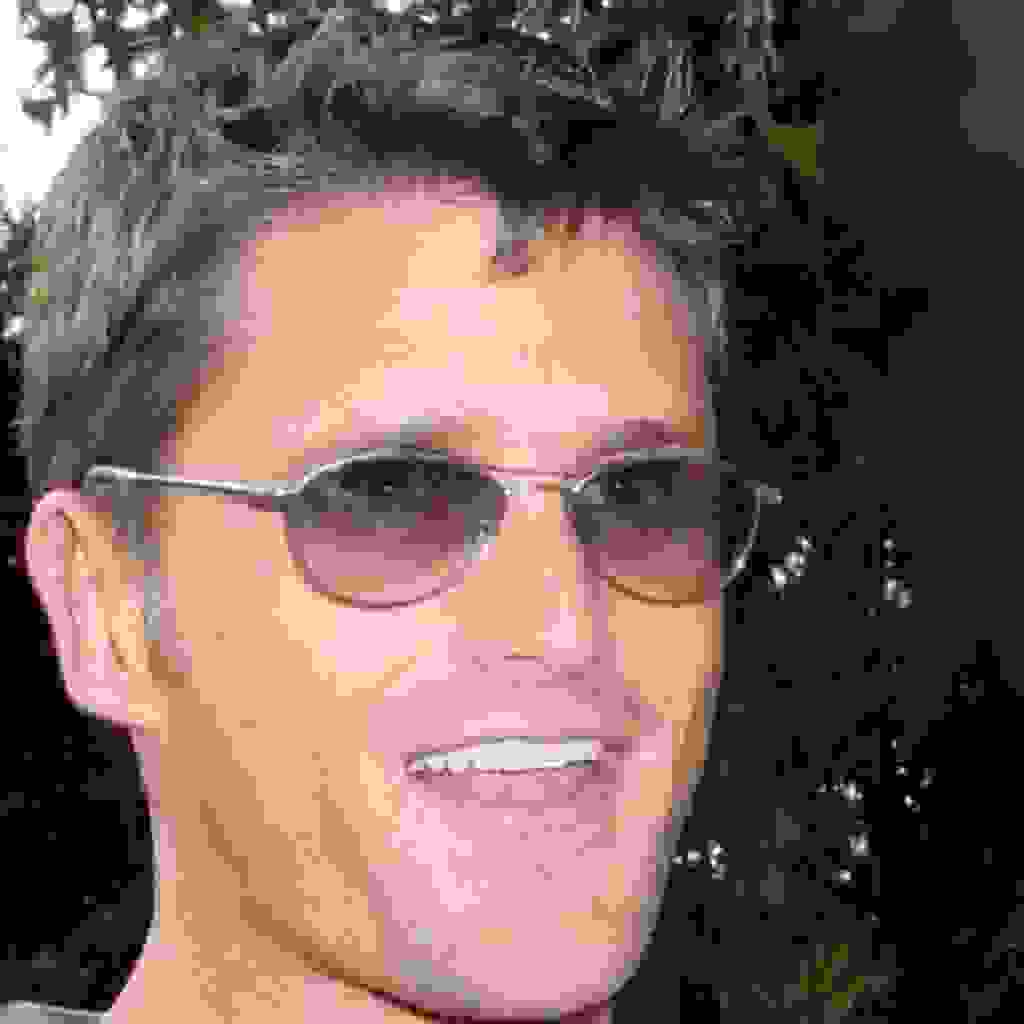}\vspace{4pt}
\includegraphics[width=1\linewidth]{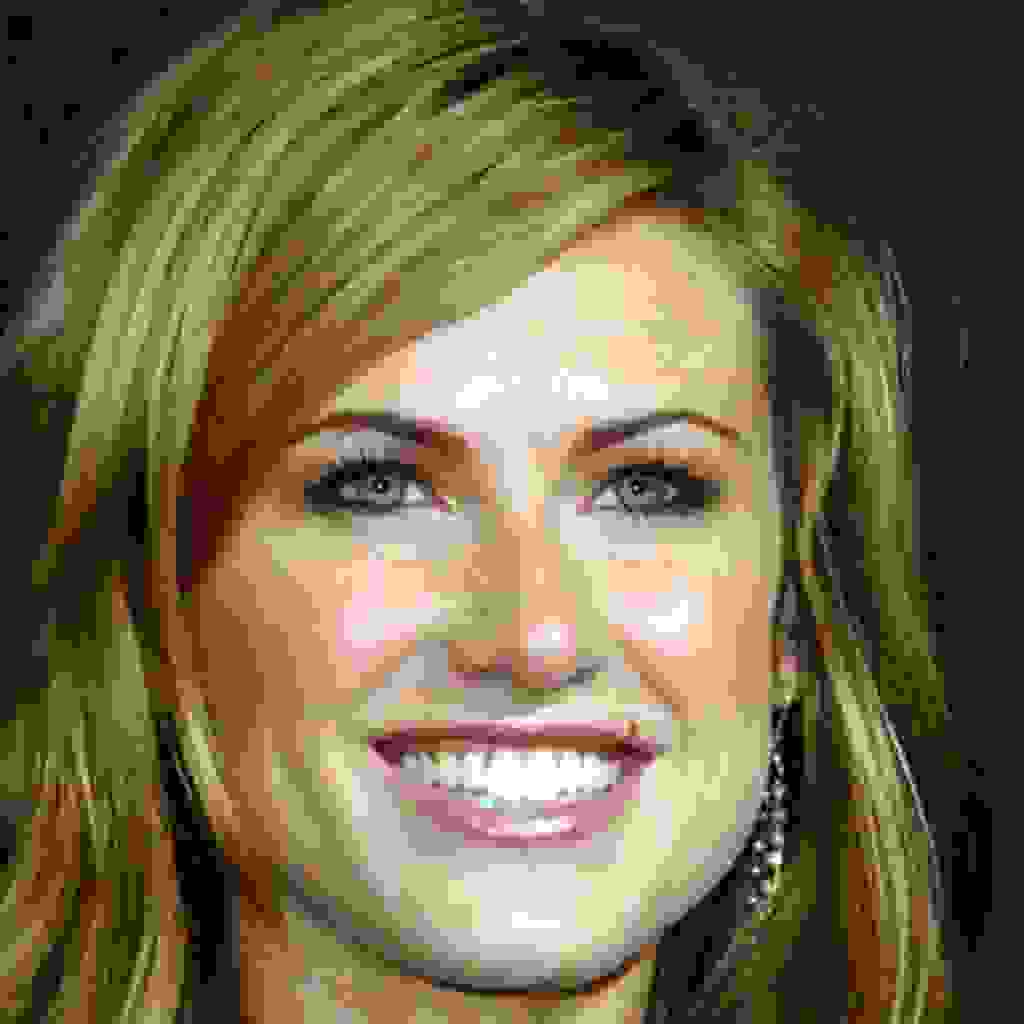}\vspace{4pt}
\includegraphics[width=1\linewidth]{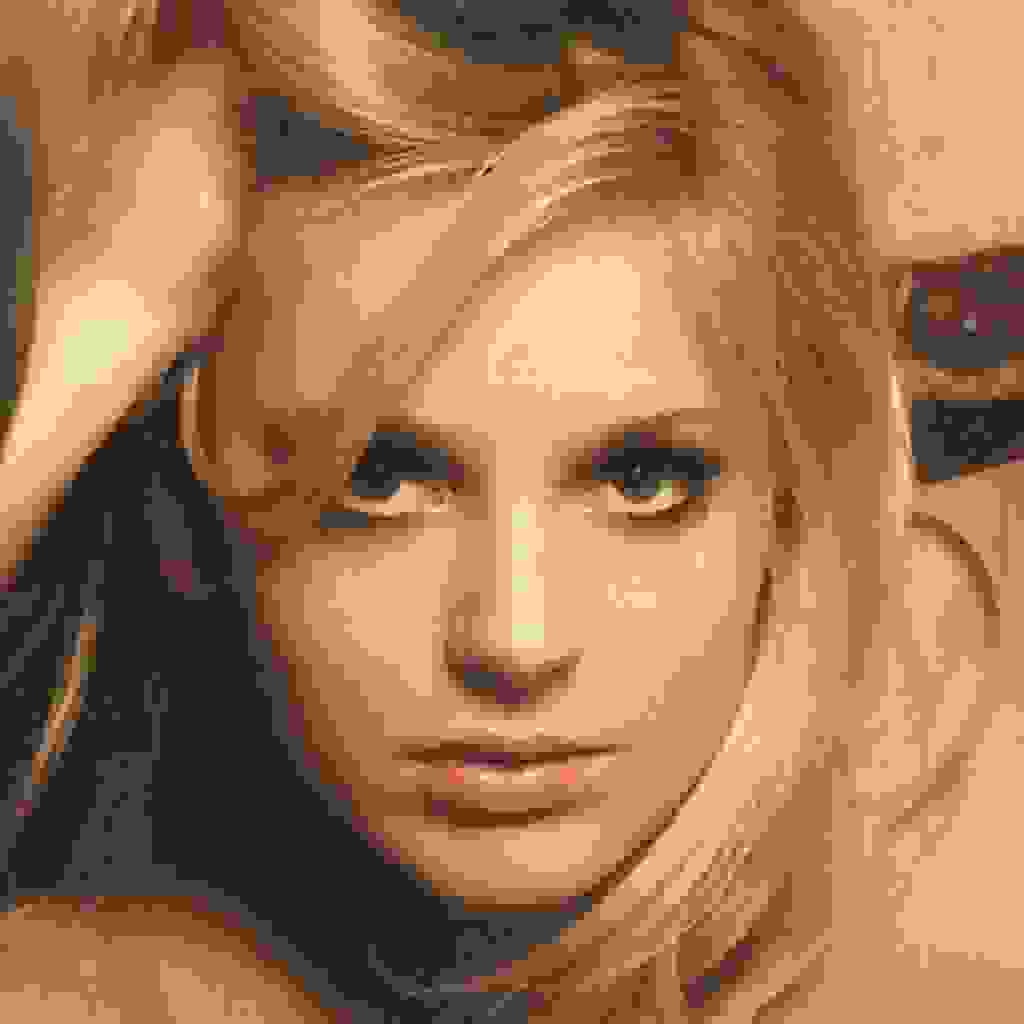}\vspace{4pt}
\includegraphics[width=1\linewidth]{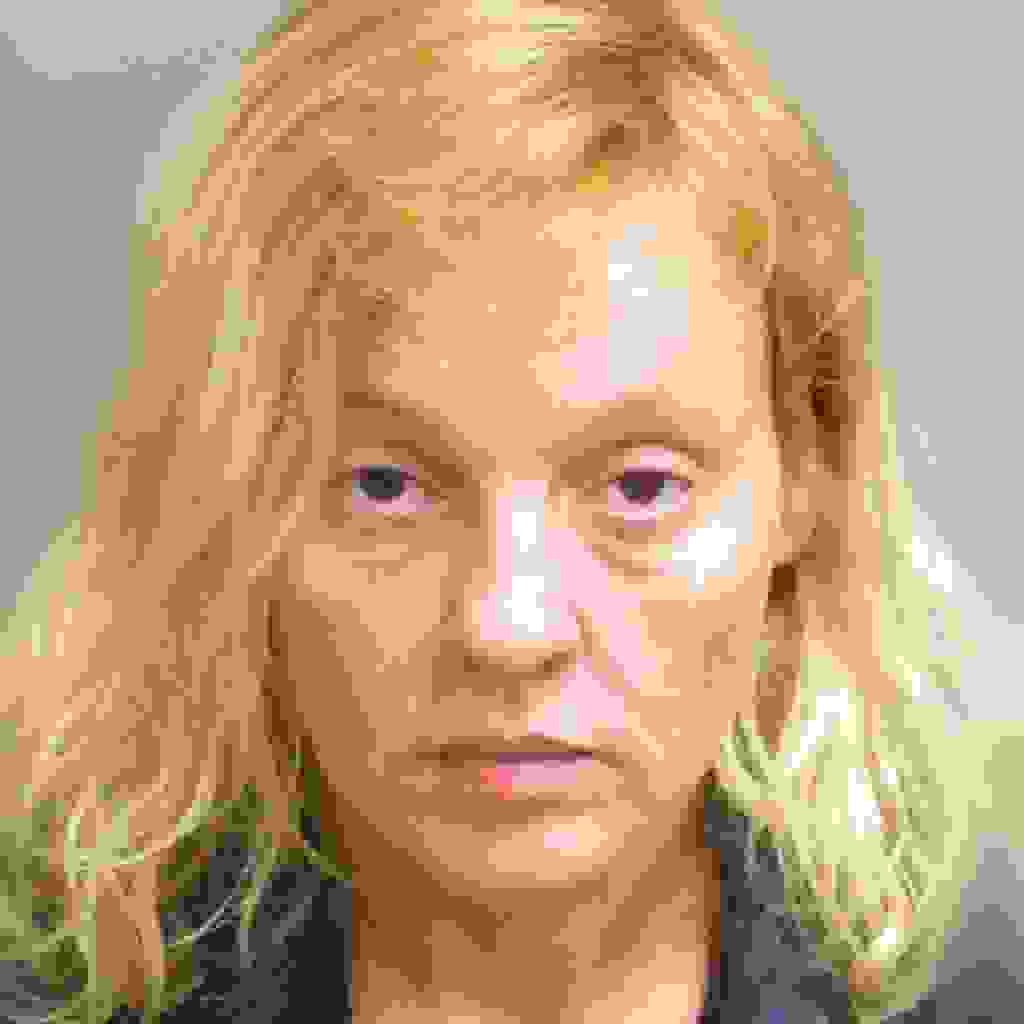}\vspace{4pt}
\end{minipage}}
\subfigure[PULSE\cite{2020PULSE}(V1)]{
\begin{minipage}[b]{0.16\linewidth}
\includegraphics[width=1\linewidth]{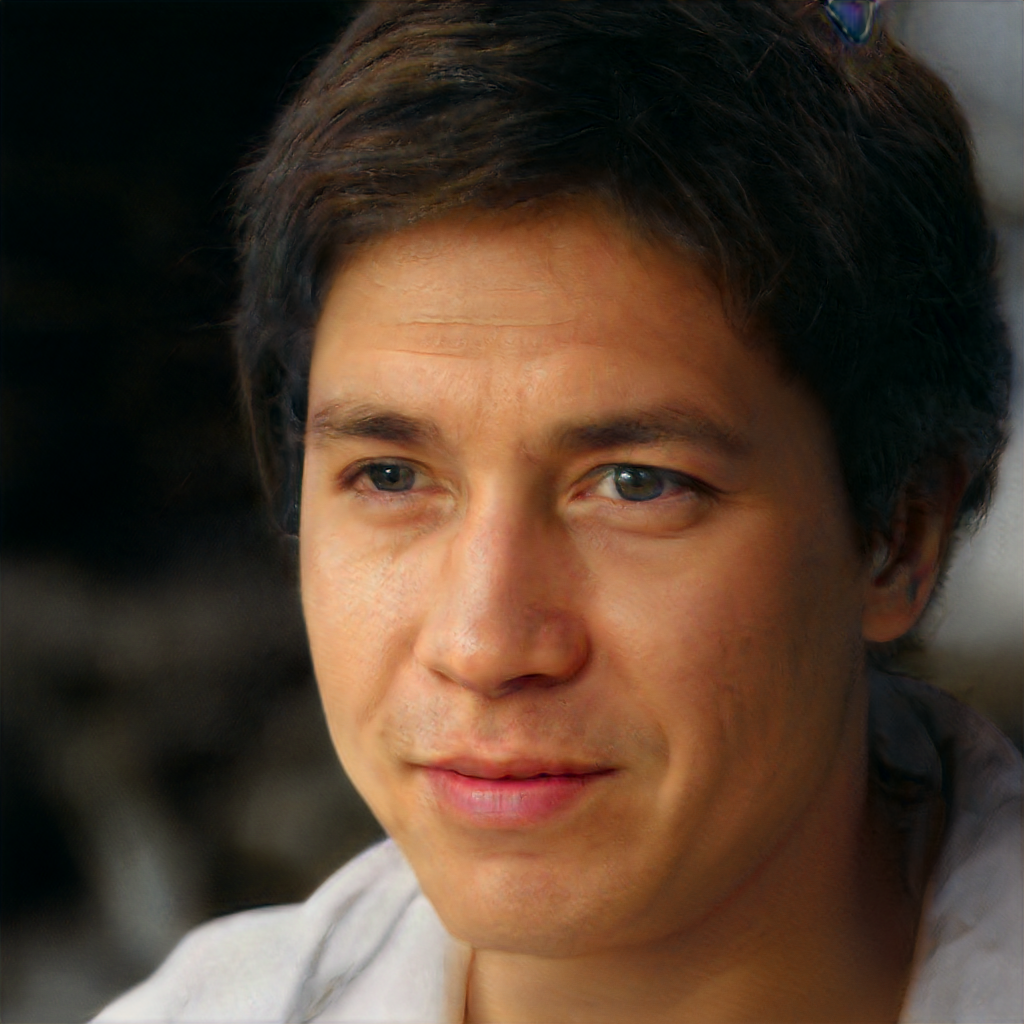}\vspace{4pt}
\includegraphics[width=1\linewidth]{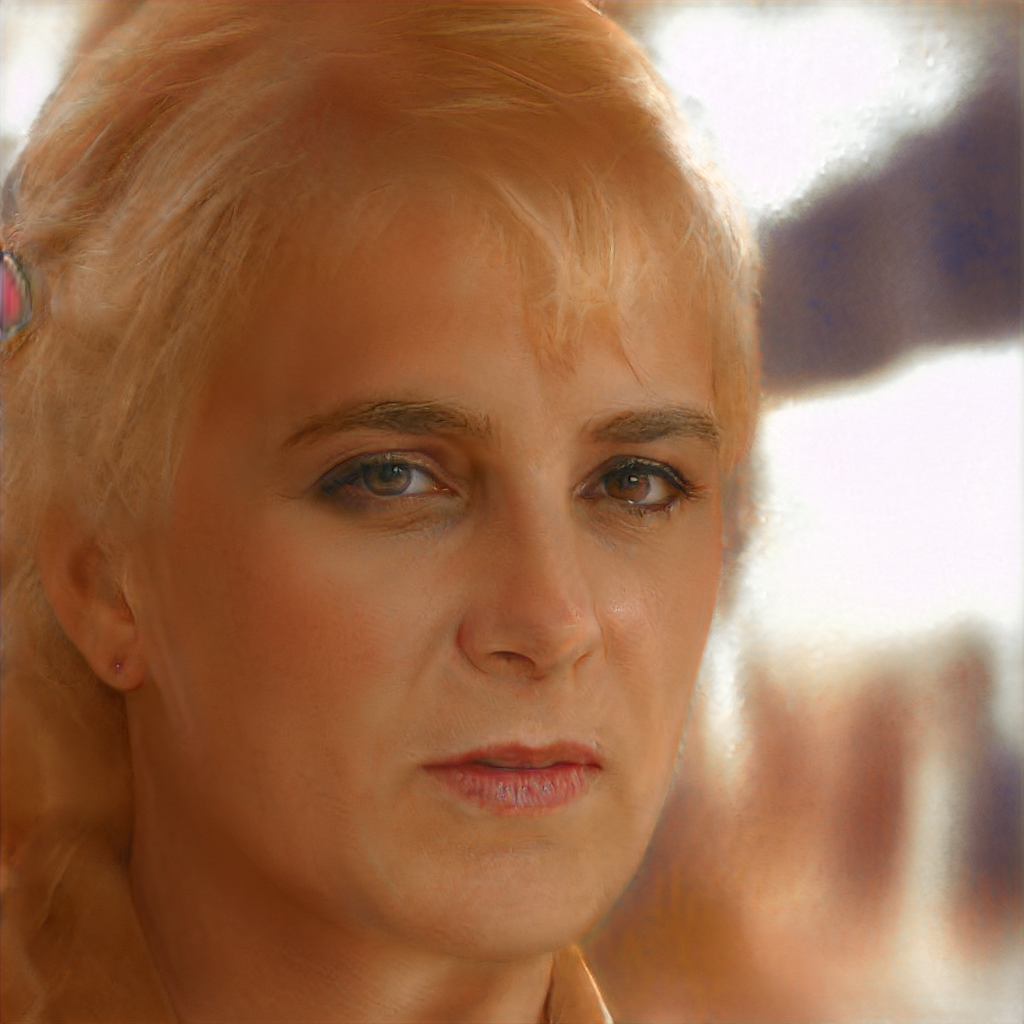}\vspace{4pt}
\includegraphics[width=1\linewidth]{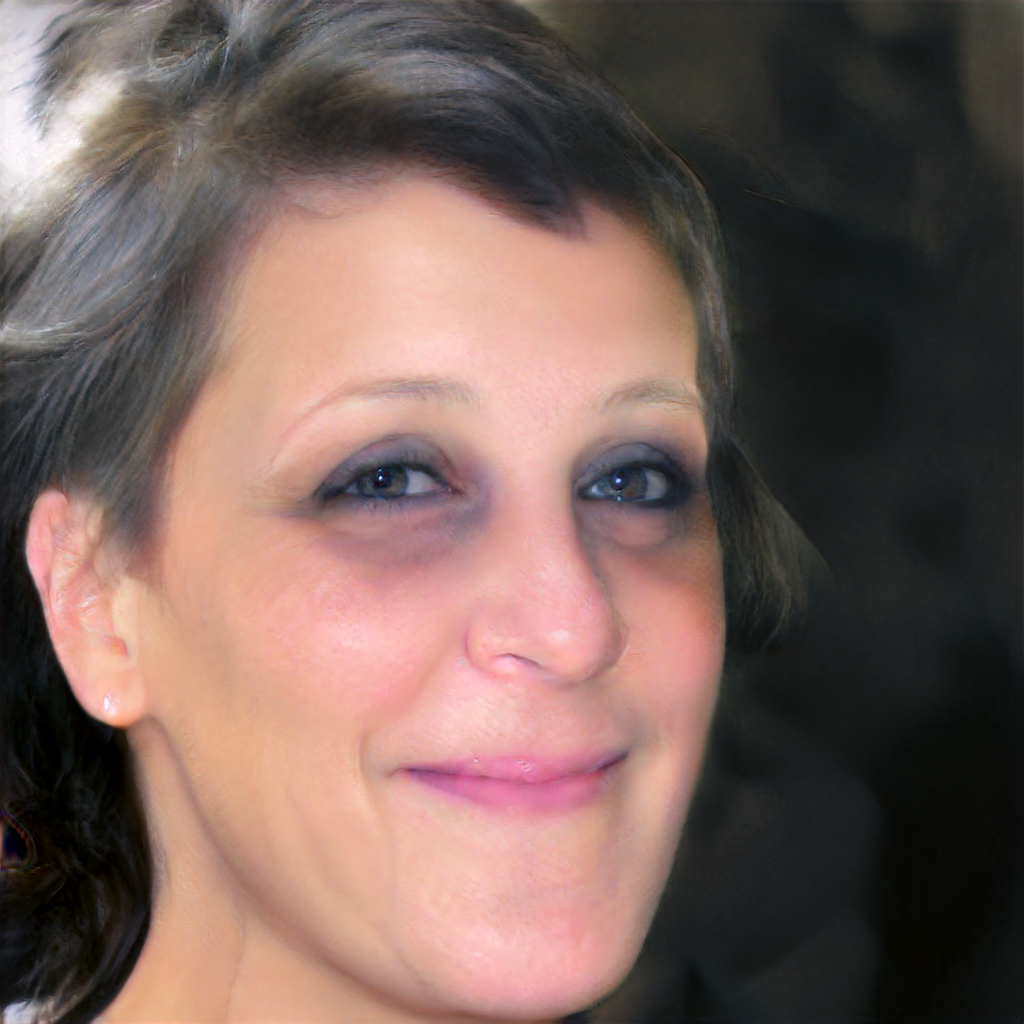}\vspace{4pt}
\includegraphics[width=1\linewidth]{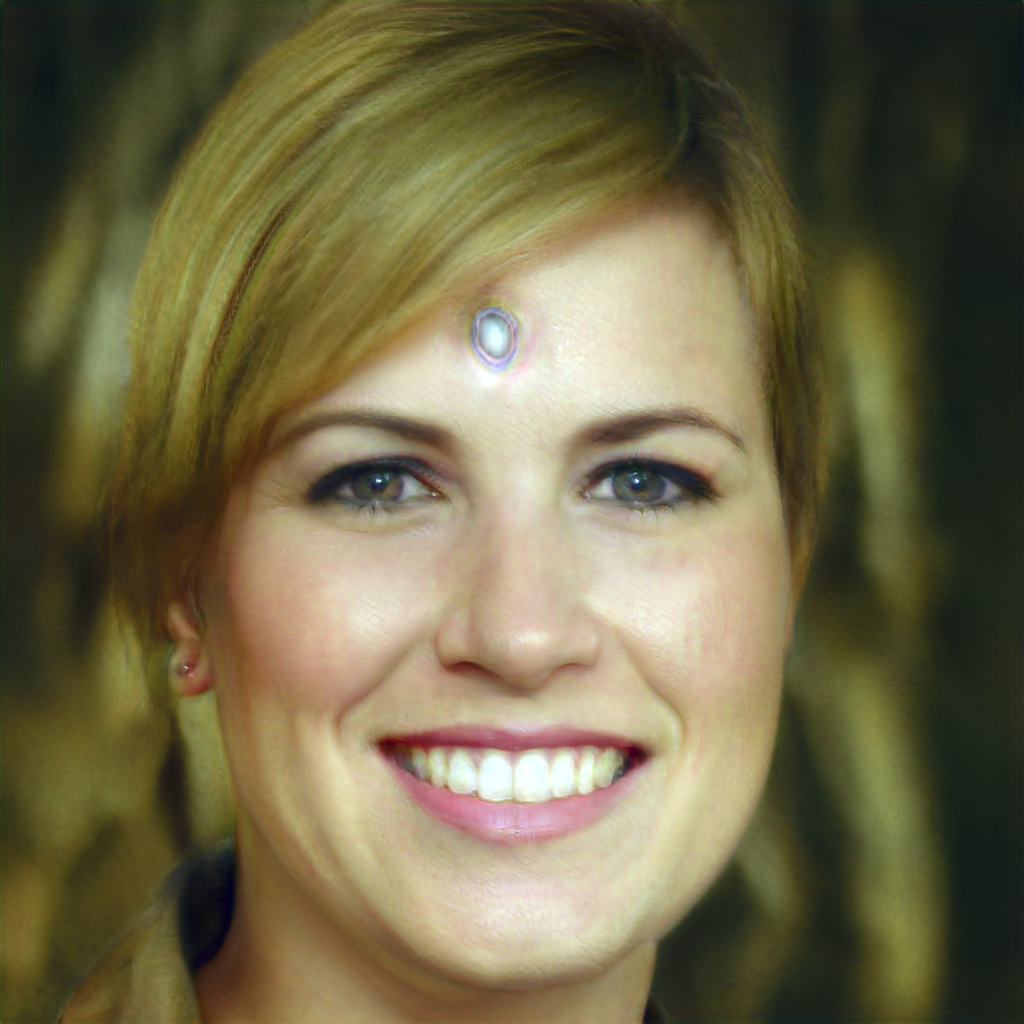}\vspace{4pt}
\includegraphics[width=1\linewidth]{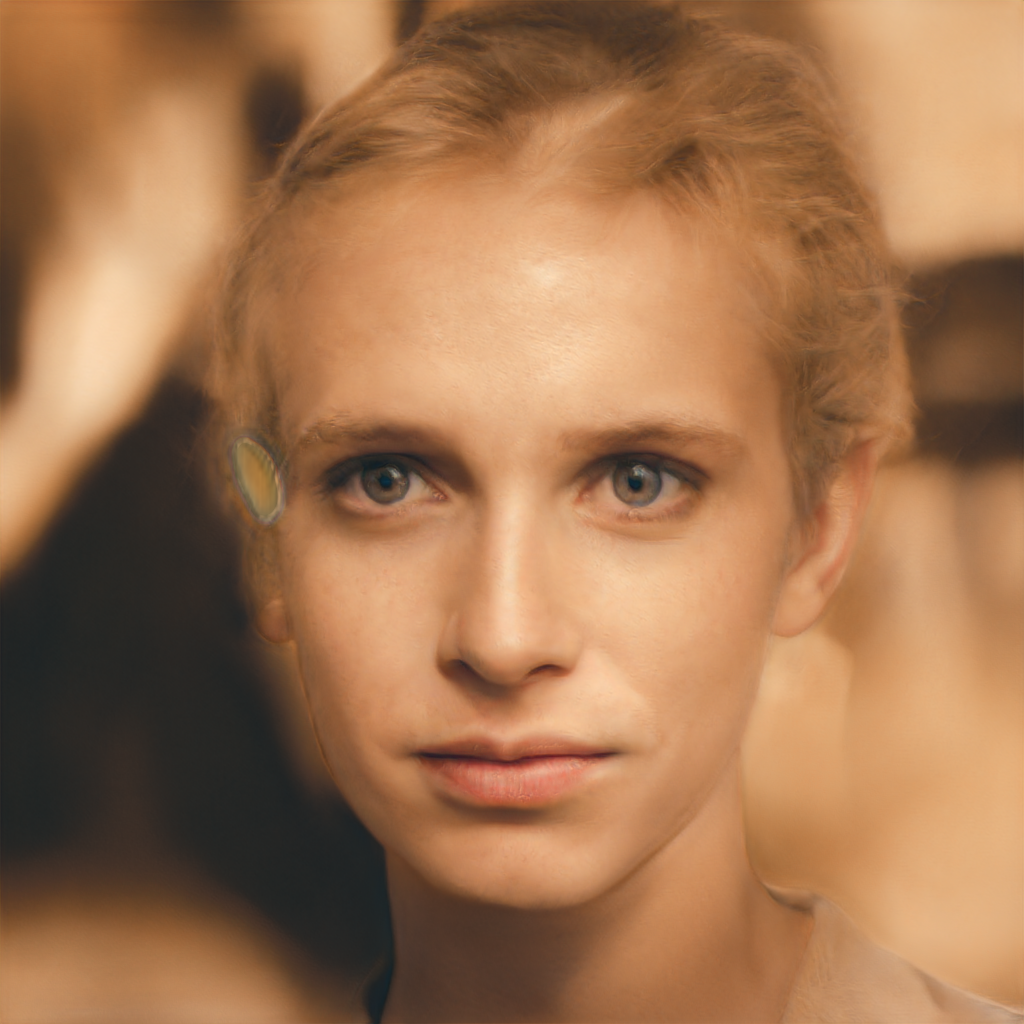}\vspace{4pt}
\includegraphics[width=1\linewidth]{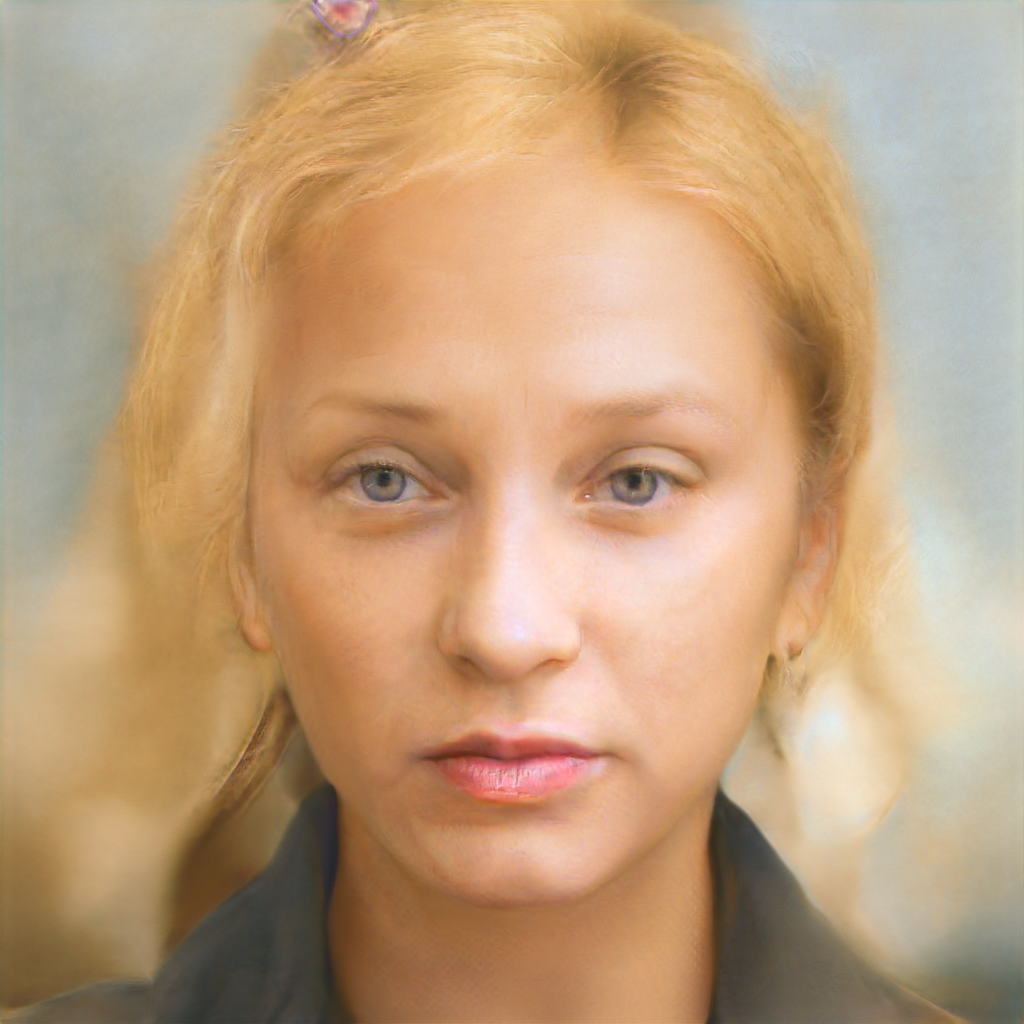}\vspace{4pt}
\end{minipage}}
\subfigure[PULSE(V2)]{
\begin{minipage}[b]{0.16\linewidth}
\includegraphics[width=1\linewidth]{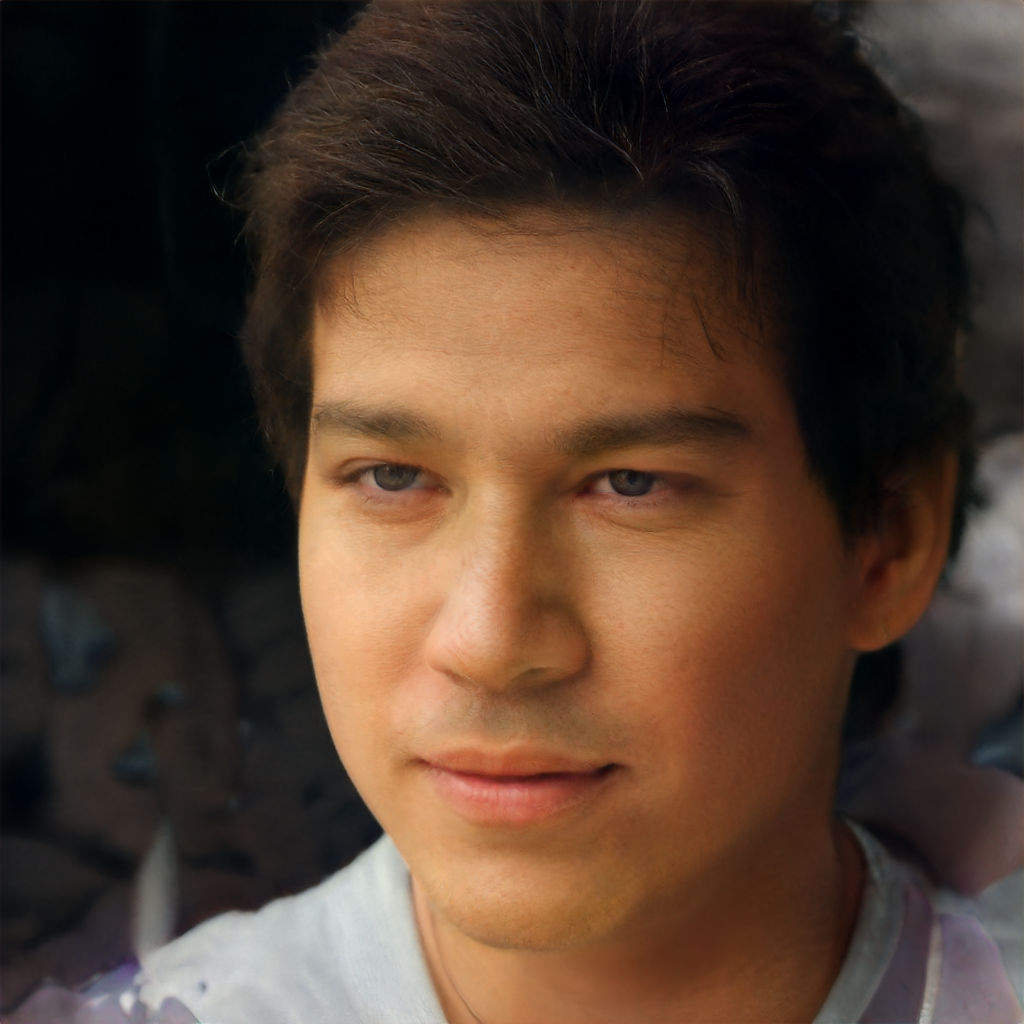}\vspace{4pt}
\includegraphics[width=1\linewidth]{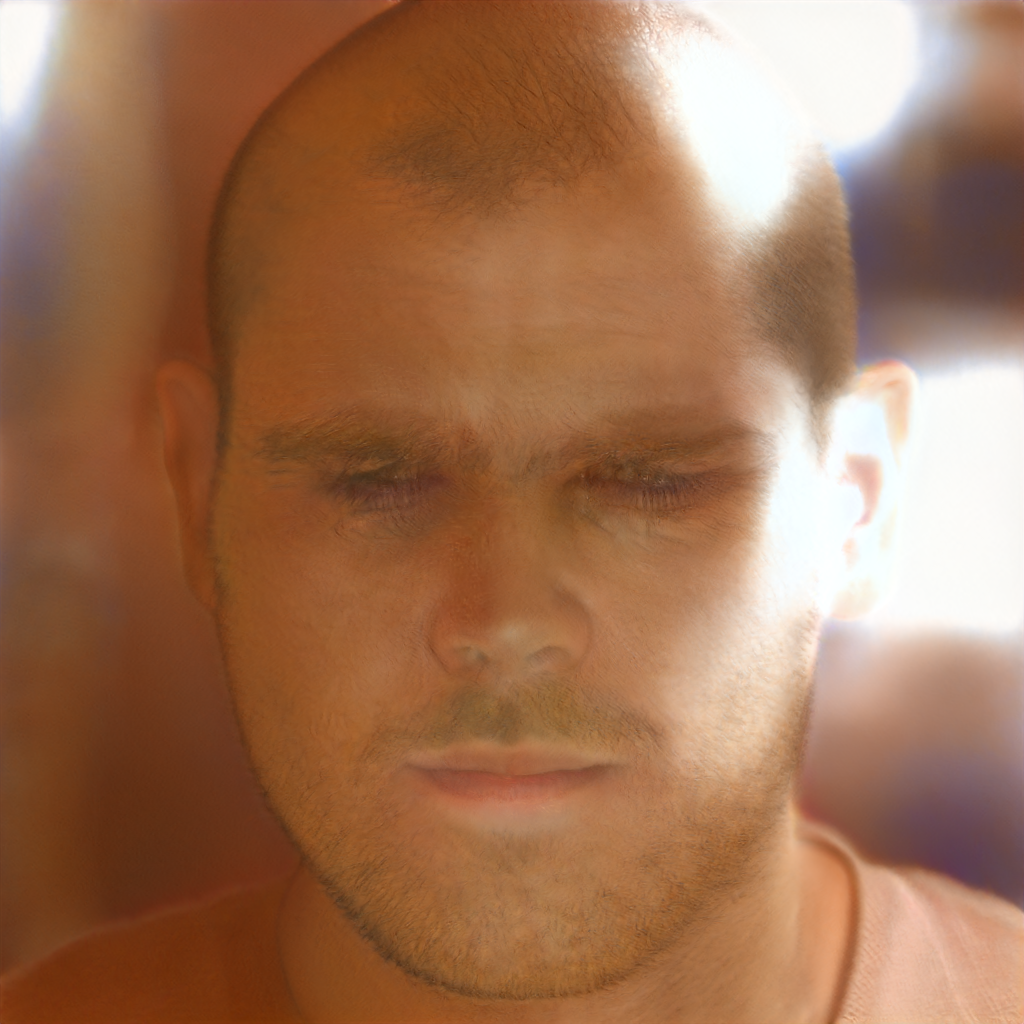}\vspace{4pt}
\includegraphics[width=1\linewidth]{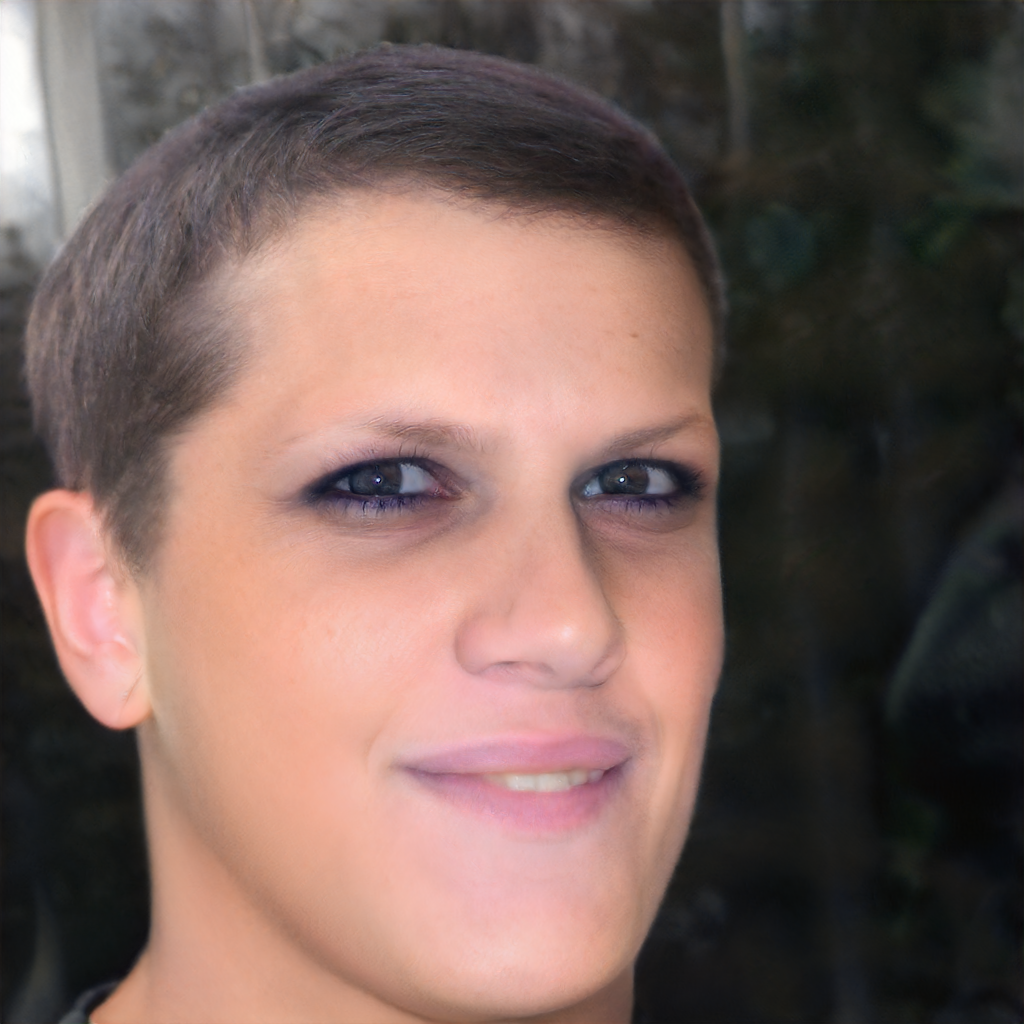}\vspace{4pt}
\includegraphics[width=1\linewidth]{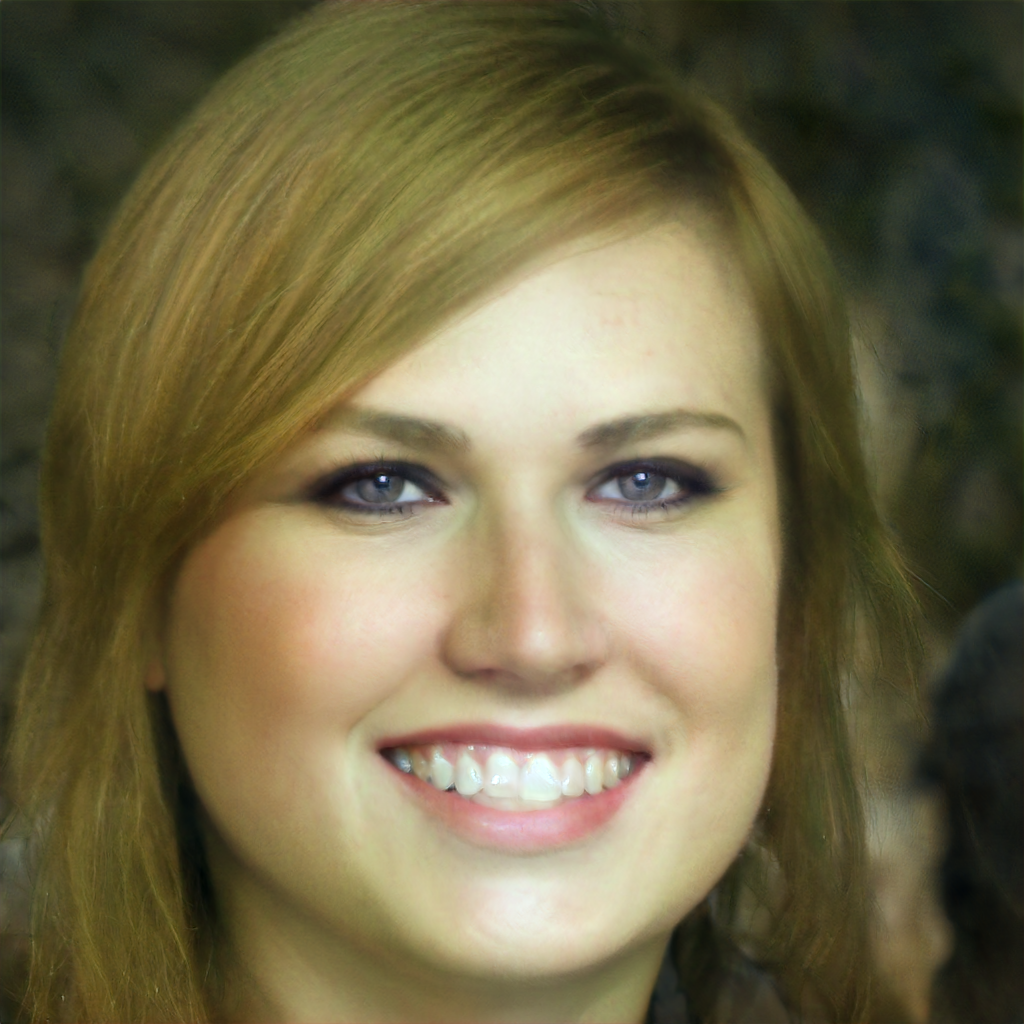}\vspace{4pt}
\includegraphics[width=1\linewidth]{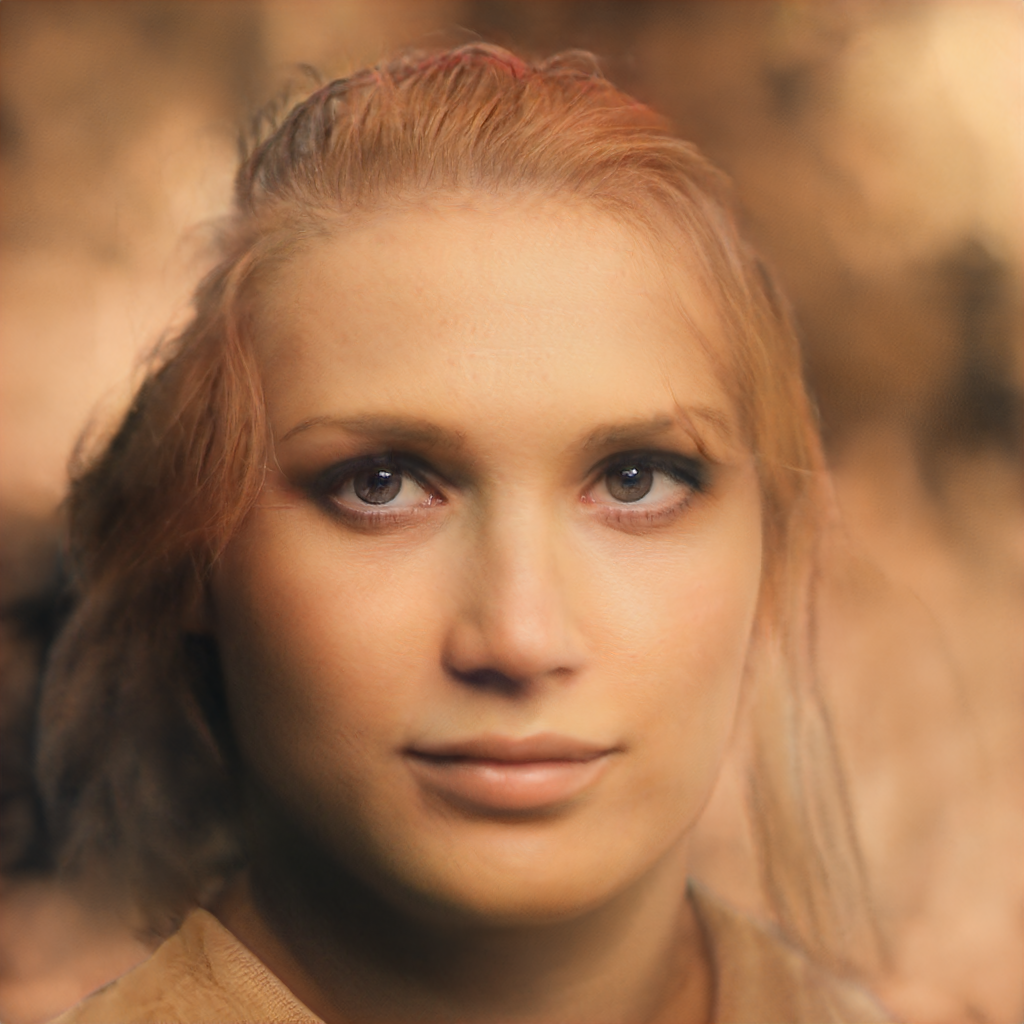}\vspace{4pt}
\includegraphics[width=1\linewidth]{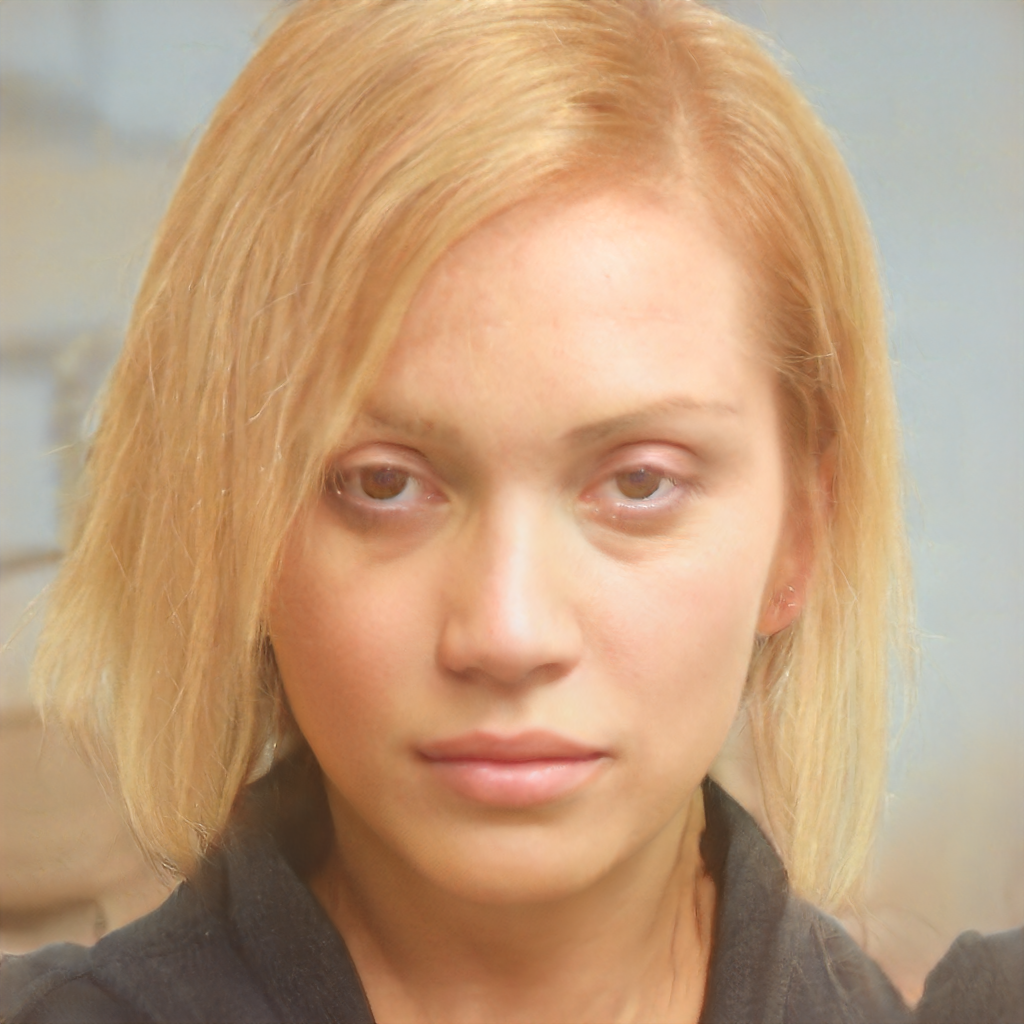}\vspace{4pt}
\end{minipage}}
\subfigure[Ours]{
\begin{minipage}[b]{0.16\linewidth}
\includegraphics[width=1\linewidth]{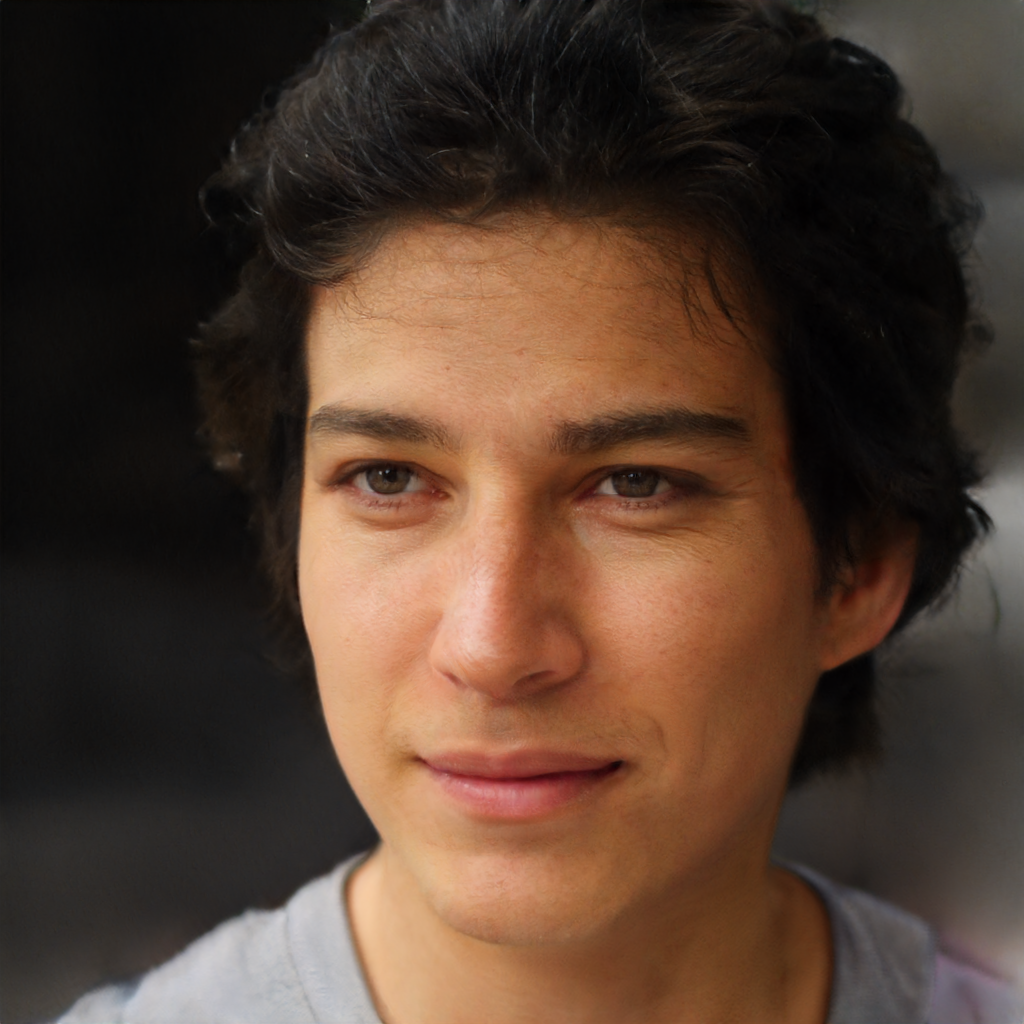}\vspace{4pt}
\includegraphics[width=1\linewidth]{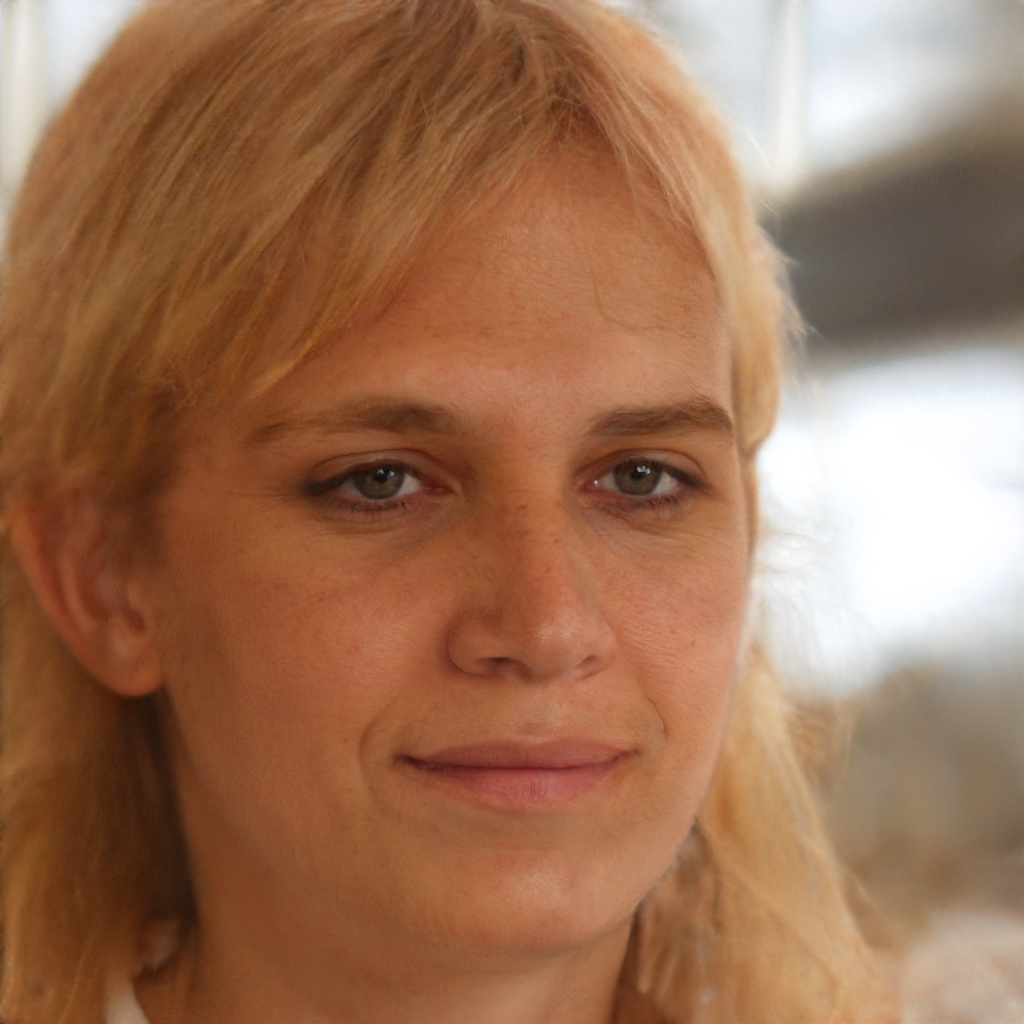}\vspace{4pt}
\includegraphics[width=1\linewidth]{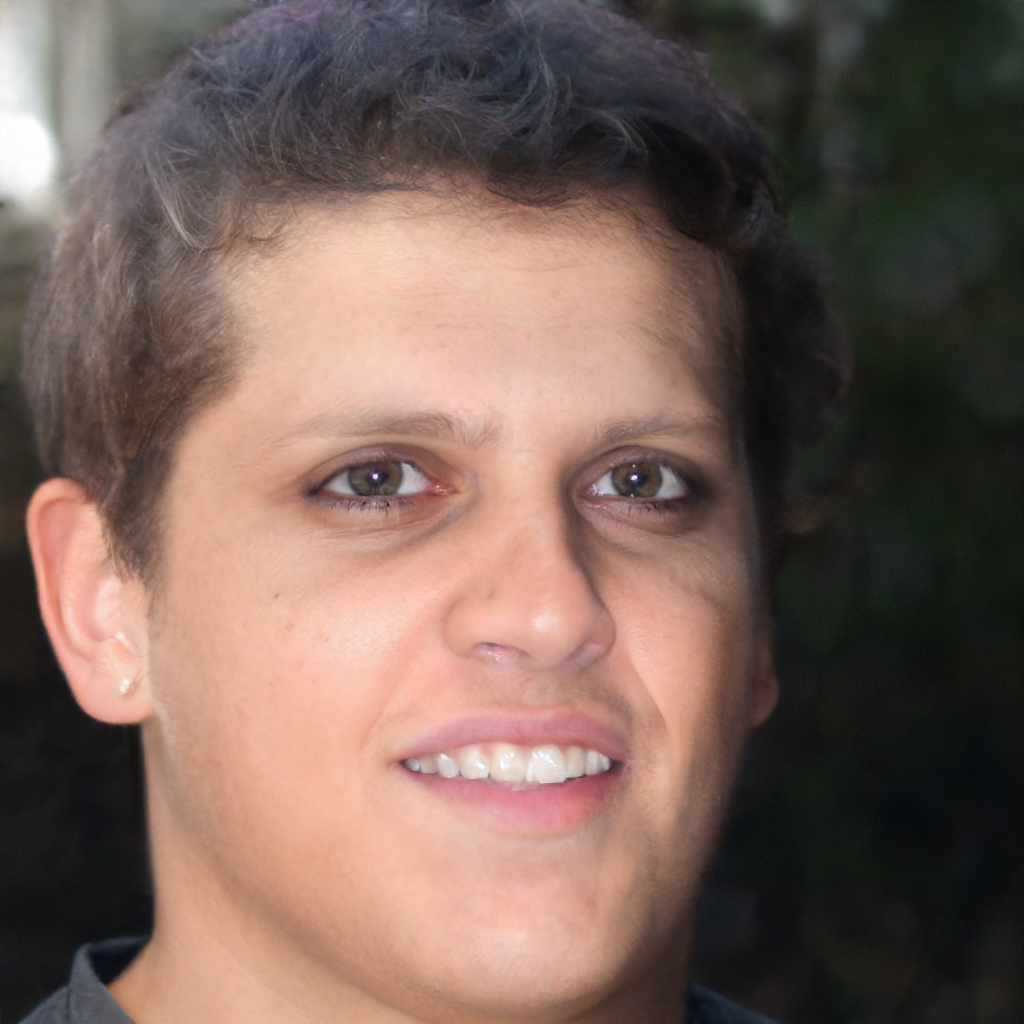}\vspace{4pt}
\includegraphics[width=1\linewidth]{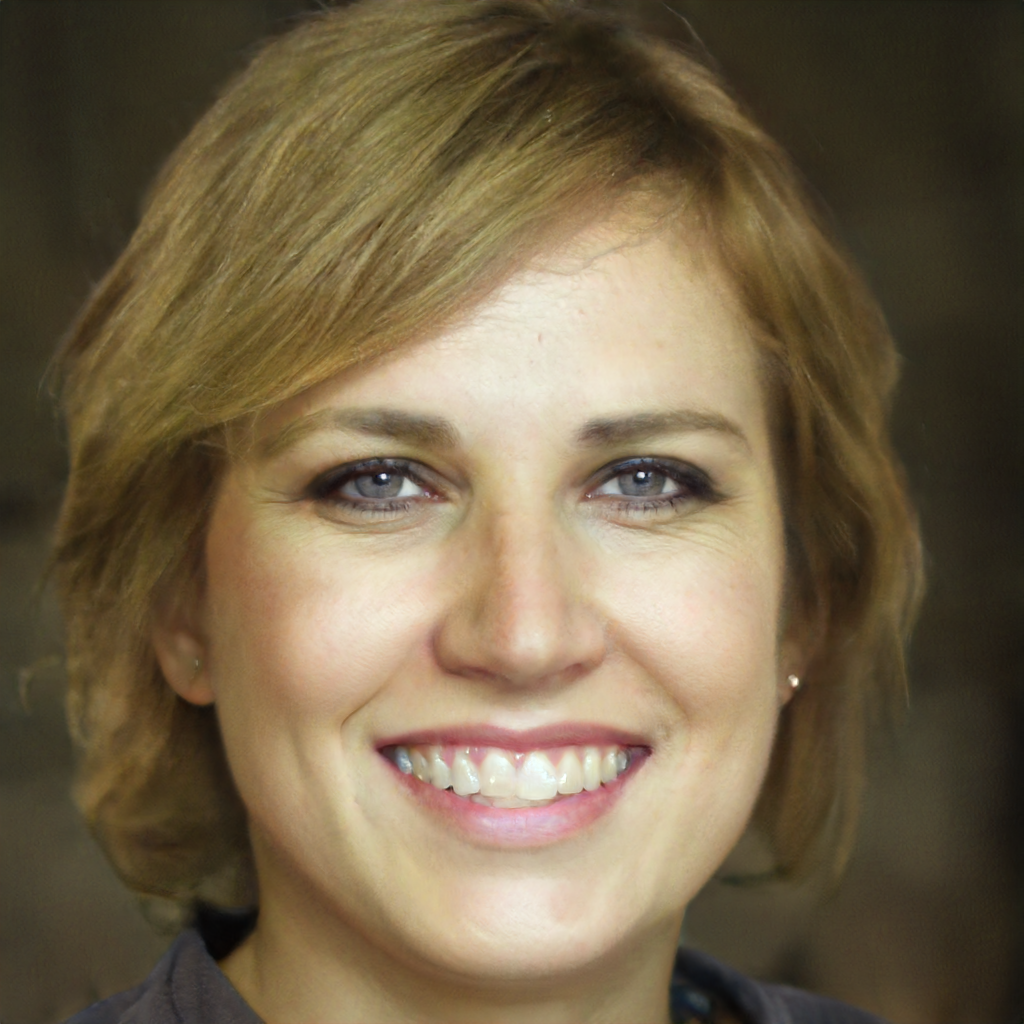}\vspace{4pt}
\includegraphics[width=1\linewidth]{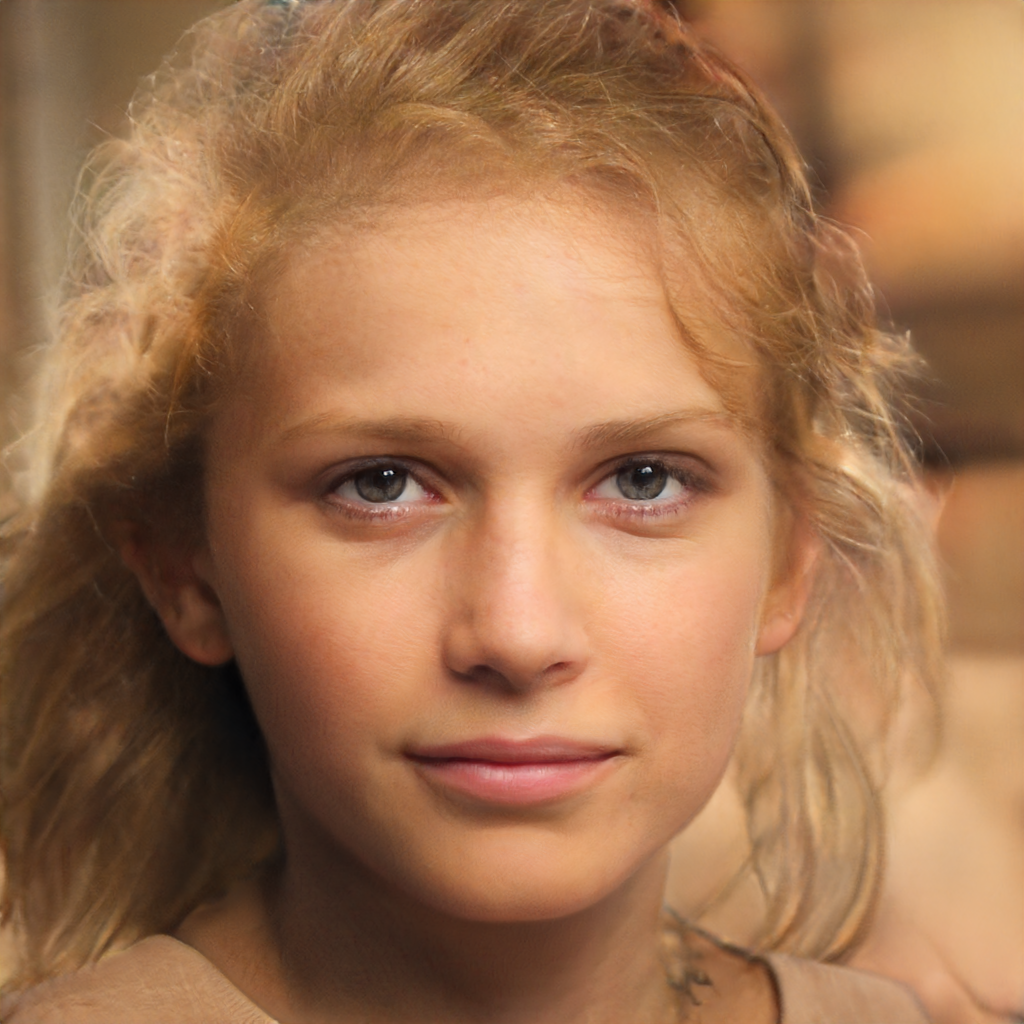}\vspace{4pt}
\includegraphics[width=1\linewidth]{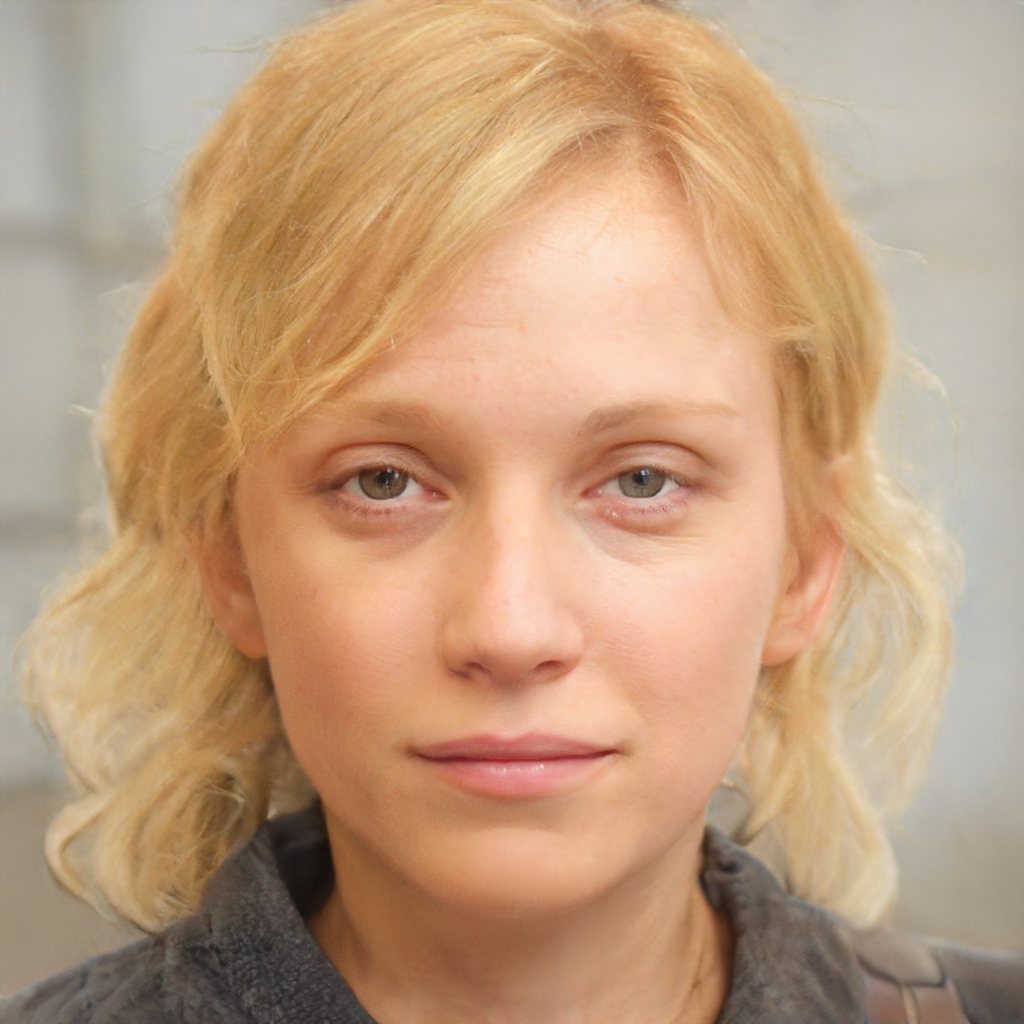}\vspace{4pt}
\end{minipage}}

\caption{Comparison of different methods. We compare with the results obtained with original StyleGAN model and the improved version on the degradation of compression. }
\label{abalation}
\end{figure*} 

\subsection{Visual quality}
We conduct extensive experiments regrading the visual quality  to demonstrate the effectiveness of the proposed method. For better comparison, we evaluate the performance from three aspects, namely the qualitative results, the user study and the no-reference image quality assessment. The experiments presented below deal with the compression degradation. The baseline PULSE method is performed with the $\ell_2$ loss, while our model incorporates the degradation model detailed above.
\subsubsection{Qualitative Image Results}
In Figure \ref{abalation} we give some examples of restored images from different methods. We display the results obtained from the PULSE method \cite{2020PULSE} with both the original StyleGAN and the improved version. We can find that the original StyleGAN contains artifacts in somewhere of the image. The improved StyleGAN successfully gets rid of such artifacts, which is in accordance with expectation. In general, our method produces the images with highest quality. 

\begin{table}
\begin{center}
\begin{tabular}{|c|c|c|} 
\hline
  & PULSE & Ours\\
\hline
 Preference &19\% & 81\% \\
\hline
\end{tabular}
\end{center}
\caption{ The preference of our method and original PULSE. By average, 81\% of the users think our results are more natural and high-quality. }
\label{MOS}
\end{table}

\subsubsection{User Study}
Following common practice, the quantitative assessment can be obtained from a user study.  
Therefore, we hereby conduct a user study to verify the effectiveness of the proposed method. Specifically, we randomly select 100 images, compress them with the JPEG, and restored with the above methods. We ask the raters to select the one that is more natural and with better visual quality.
The preference of our method is shown in Table \ref{MOS}. 
The averaged result shows that 81\% of the users think our results are more natural and high-quality compared with PULSE.

\subsubsection{No-reference Image Quality Assessment}

Besides the subjective score, we also give comparsion between above methods leveraging recent advances on no-reference image quality  assessment (NR-IQA).
We choose  the state-of-the-art model RankIQA \cite{RankIQA} to accomplish this task. Specifically, we adopt the model pretrained in the LIVE \cite{LIVE} dataset. 
For a particular image, we adopt two strategies computing the quality score. Firstly, we randomly crop 100 patches with size 224x224 and report the average and median scores as the quality index. 
Secondly, we resize the images to $256\times256$ and re-run the first method.
The second strategy deals with the unmatched size on which the RankIQA is trained, and ensures the model has the picture of the whole image.
The score ranges from 0 to 100 in LIVE dataset and smaller index indicates better visual quality. 
We conduct extensive ablation study to further demonstrate the improvements of different modules. The result is shown in Table \ref{Score}. We find the proposed method achieve better results on both the two strategies.





{\small
\bibliographystyle{unsrt} 
\bibliography{egbib}
}

\end{document}